\documentclass[usegraphicx,useAMS,usenatbib]{mn2e}
\usepackage{graphicx}
\usepackage{ulem}
\usepackage{amssymb}

\def\name{PHOX}
\def\leiaD{CL1}
\def\leiaB{CL2}

\def\fig{Fig.\,}
\def\eq{Eq.\,}
\def\sec{Section~}
\def\tab{Table\,}

\def\mfive{M_{500}}
\def\rfive{R_{500}}

\def\xspec{{\tt XSPEC}}
\def\xissim{{\tt XISSIM}}
\def\arfgen{{\tt XISSIMARFGEN}}
\def\apec{{\tt APEC}}
\def\bapec{{\tt BAPEC}}
\def\mekal{{\tt MEKAL}}
\def\vapec{{\tt VAPEC}}
\def\bvapec{{\tt BVAPEC}}
\def\vmekal{{\tt VMEKAL}}
\def\wabs{{\tt WABS}}
\def\zsun{\rm{Z_{\odot}}}

\def\msun{\rm{M_{\odot}}}
\def\cm{\rm{cm}}

\def\mpc{\rm{Mpc}}
\def\kpc{\rm{kpc}}
\def\kev{\rm{keV}}

\def\lcdm{$\Lambda$CDM }
\def\om{\Omega_0}
\def\omb{\Omega_b}

\title[\name{}: a novel X--ray photon simulator]{Observing simulated galaxy clusters with \name{}: a novel X--ray photon simulator}

\author[V. Biffi et al.]{V. Biffi$^{1,2}$\thanks{E--mail:
biffi@mpa-garching.mpg.de}, K. Dolag$^{1,3}$, H. B\"ohringer$^2$ and G. Lemson$^{1}$\\
$^{1}$Max--Planck--Institut f\"ur Astrophysik, Karl--Schwarzschild--Strasse 1, 85748 Garching bei M\"unchen, Germany\\
$^{2}$Max--Planck--Institut f\"ur extraterrestrische Physik, Giessenbachstrasse 1, 85748 Garching bei M\"unchen, Germany\\
$^{3}$University Observatory Munich, Scheinerstrasse 1, 81679 M\"unchen, Germany}

\begin{document}

\date{Accepted ... Received ... ; ...}
\pagerange{\pageref{firstpage}--\pageref{lastpage}} \pubyear{...}
\maketitle
\label{firstpage}


\begin{abstract}
We present a novel, virtual X--ray observatory designed to obtain synthetic observations from hydro--numerical simulations, named \name{}. In particular, we provide a description of the code constituting the photon simulator and of the new approach implemented. We apply \name{} to simulated galaxy clusters in order to demonstrate its capabilities. In fact, X--ray observations of clusters of galaxies continue to provide us with an increasingly detailed picture of their structure and of the underlying physical phenomena governing the gaseous component, which dominates their baryonic content. Therefore, it is fundamental to find the most direct and faithful way to compare such observational data with hydrodynamical simulations of cluster--like objects, which can currently include various complex physical processes.
Here, we present and analyse synthetic {\it Suzaku} observations of two cluster--size haloes obtained by processing with \name{} the hydrodynamical simulation of the large--scale, filament--like region in which they reside. 
Taking advantage of the simulated data, we test the results inferred from the X--ray analysis of the mock observations against the underlying, known solution.
Remarkably, we are able to recover the theoretical temperature distribution of the two haloes by means of the multi--temperature fitting of the synthetic spectra. Moreover, the shapes of the reconstructed distributions allow us to trace the different thermal structure that distinguishes the dynamical state of the two haloes.
\end{abstract}




\section{INTRODUCTION}\label{sec:intro}

As optimal laboratories for both cosmology and astrophysics, galaxy clusters have been thoroughly investigated with large surveys as well as via dedicated single--object observations. In this respect, X--ray measurements still represent one of the best ways to fully study their structure, traced by the hot plasma filling their potential wells. In particular, the most important and difficult quantity to infer is the total gravitating mass, which can be determined from measurements of the intra--cluster medium (ICM) temperature and density profiles, assuming hydrostatic equilibrium to hold. 
So far, studies of the X--ray--emitting ICM have been mainly limited to the innermost region, where the investigation of the baryonic physics is observationally less challenging than in the outskirts. Although very difficult to map in the X rays, a characterization of the outer region of clusters is in fact crucial to understand the formation and evolution of this structures and to use them as cosmological probes, since it encloses a significant fraction of the cluster volume, where the properties of the accreting gas and of the dark matter halo are still uncertain \cite[e.g., see][]{ettori2011}.
Only very recently, X--ray observations performed with the Japanese satellite {\it Suzaku} have reached the virial radius for a few clusters of galaxies \cite[e.g.][]{fujita2008,bautz2009,george2009,reiprich2009,hoshino2010,kawaharada2010,simionescu2011,akamatsu2011}, promisingly taking a step forward on the temperature profile debate and indirectly opening new perspectives in controlling the biases on cluster--based cosmological investigations. These results are just one example of the improvements by which X--ray observations continue to provide us with an increasingly detailed picture of galaxy clusters, for which a clear interpretation of the underlying physical processes is very challenging.

On the theoretical side, many interesting issues on cluster formation and evolution are addressed by means of hydro--numerical simulations, which represent a powerful tool to investigate in detail the nature of these complicated astrophysical objects \cite[e.g.][for a recent review]{borgani2009}. Simulations of large cosmological boxes that include several massive cluster--like haloes can now be performed incorporating not only the dominating dark matter component but also baryonic matter, in the form of gas-- and star--like particles (or cells), governed by hydrodynamical processes. With particular concern for galaxy clusters, current simulations are not only able to account for basic gas hydrodynamics, but implement also more complicated models for 
star formation from multi--phase medium \cite[e.g.][]{katz1992,katz1996,springel2003,marri2003} and 
thermal or kinetic feedback from supernovae--driven winds \cite[][]{navarro1993,scannapieco2006,dallavecchia2008},
chemical enrichment, metal and molecule cooling \cite[e.g.][]{mosconi2001,yoshida2003,tornatore2004,tornatore2007,scannapieco2005,maio2007,maio2010},
thermal conduction \cite[][]{cleary1999,jubelgas2004,ruszkowski2010}, 
AGN feedback \cite[e.g.][]{springel2005b,dimatteo2005,sijacki2006,sijacki2007,sijacki2008,puchwein2008,fabjan2010,dubois2010,teyssier2011}, 
cosmic rays \cite[][]{pfrommer2007,jubelgas2008} and
magnetic fields \cite[][]{phillips1985,dolag1999,brueggen2005,price2005,dolag2009}, 
to name the most important effects.

The ideal achievement would be the combination of both these sources of information, by directly comparing simulated clusters to X--ray observations of real objects. Devoted to this goal, sophisticated numerical codes, such as X--MAS/X--MAS2 \cite[][]{gardini2004,rasia2008} and, more recently, XIM \cite[][]{xim2009,heinz2010}, have been developed in the last years in order to turn hydro--simulation outputs into mock images with a given X--ray telescope. 
In particular, the X--MAS virtual telescope is explicitly designed to process outputs obtained from SPH codes like GADGET, whereas XIM is particularly dedicated to grid--based hydrodynamical simulations, e.g. performed with the AMR code FLASH \cite[][]{fryxell2000}, which makes them fundamentally complementary.
Similarly, both X--MAS and XIM use a plasma thermal emission code by which they calculate the projected emission associated to the gas component in the simulation. The computational effort, required to calculate this, is usually reduced by interpolating among the model spectra externally stored in a library of templates.
Essentially, the emissivity integrated along the chosen line of sight is calculated in terms of flux, depending on the properties of all the gas at a given projected position. 
Lastly, the convolution with instrumental response is performed and the final photon event file is generated.
Such virtual telescopes represent an important step forward with respect to visualization tools used to produce surface brightness maps from simulation outputs, since they properly calculate the emission associated to the gas component accounting for the structure of the simulated source along the line of sight. Moreover, they are able to include accurate calculations required to obtain realistic mock X--ray observations, for instance convolving the simulated spectra with any given instrument response and telescope PSF.
In fact, it is vital that the synthetic observations produced by such simulators are as much as possible similar to the standards commonly used by observers, so that the comparison can be the most faithful.
As such, this advanced approach easily leads to a challenging increase of the parameters to handle in order to realistically model the plasma emission and the detailed three--dimensional structure of the simulations.

Analogous in the scope, the virtual telescope presented in this paper, \name{}, is also dedicated to convert hydro--numerical outputs into mock X--ray observations. 
However, the novelty of our simulator lies in the method adopted, by which the spectral emission calculated singularly for each gas element in the simulation is immediately converted into a discrete sample of photons, collected and stored before projecting along any line of sight and convolving with any desired instrument. 
With this strategy we are able to
significantly gain in computational effort, 
since it requires to process the original simulation only once,
independently of the specific synthetic observation to be performed afterwards. 
Furthermore, the guarantee for high spatial and spectral resolution, preserved without dramatically increasing the computational cost,
offers the possibility to anticipate the observational achievements of upcoming X--ray missions, such as {\it IXO/ATHENA},
in which the high--resolution spectroscopy will allow us to explore the intrinsic structure of galaxy clusters through the study of their spectral features. In fact, it has been suggested in the last decade that the diagnostics of the broadening of heavy--ion emission lines in highly--resolved X--ray spectra of galaxy clusters could actually provide valuable information about the underlying structure of the ICM and its velocity field (see \cite{pawl2005}, for a preliminary, interesting work in this field), useful for the dynamical classification of these objects.  
The expectations for such line diagnostics are related in particular to the most prominent emission line in X--ray spectra, 
namely the $\sim 6.7 \kev$ line from helium--like iron. In fact, the large atomic mass of the FeXXV ion significantly reduces the thermal line broadening, so that the line width turns out to be definitely more sensitive to turbulent or bulk gas motions \cite[][]{inogamov2003,sunyaev2003}.
This kind of studies can eventually help constraining the detectability of non--thermal motions that are likely to establish in the ICM \cite[e.g.][]{rebusco2008} and can compromise X--ray mass measurements \cite[][]{rasia2006,fang2009,lau2009,biffi2011}.

The paper is structured as follows: \sec\ref{sec:method} is devoted to the description of the method implemented in the photon simulator and the fundamental units of the code. In \sec\ref{sec:application} we apply \name{} to the hydrodynamical simulation of a filament--like structure, describing in detail each phase, from the photon generation till the mock {\it Suzaku} observation of two massive galaxy clusters residing in the filament and the spectral analysis of the synthetic spectra. Results about the recovering of the ICM emission measure distribution of the two cluster--like haloes from the multi--temperature fitting are presented in \sec\ref{sec:results}. Finally, we discuss the results obtained, summarise and draw our conclusions in \sec\ref{sec:conclusion}.

%
\section{\name{}: THE TECHNIQUE}\label{sec:method}
\begin{figure}
	\begin{center}
            \includegraphics[width=0.8\textwidth]{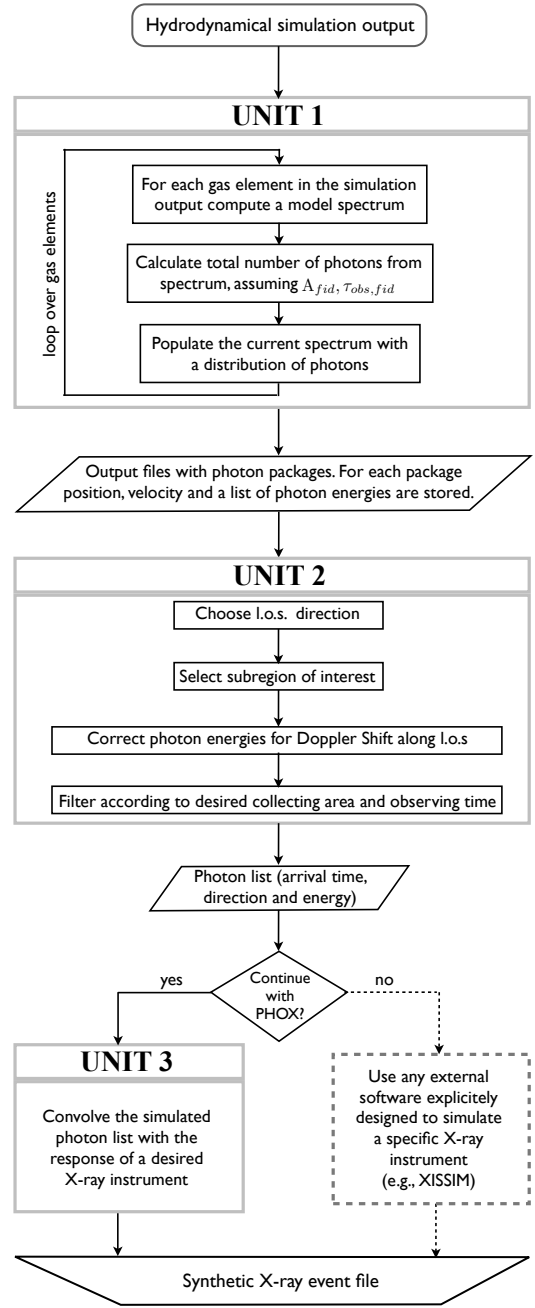}
            \caption{Flow chart illustrating the main units of \name{}. The dashed box represents the possibility to couple \name{} with any external software specifically designed to simulate observations with a certain X--ray telescope.}
		\end{center}
            \label{fig:code}
        \end{figure}

The novelty of our photon simulator arises from the method implemented. In this Section we describe the three main parts of the code that constitute the virtual telescope, \name{}\footnote{The source code of the presented version of \name{} is made available by request to the authors (see http://www.mpa-garching.mpg.de/~kdolag/Phox/). We also plan to make the code available directly for anonymous download in the near future.}, 
for which a schematic overview is also presented in \fig\ref{fig:code}. 
\name{} consists of a C code, which makes use of the publicly available X--ray package \xspec{} \cite[see][]{xspec1996} 
to calculate model emission spectra, and of the GSL\footnote{See http://www.gnu.org/s/gsl/.} 
and CFITSIO\footnote{See http://heasarc.gsfc.nasa.gov/fitsio/.} 
external, auxiliary libraries.
 
With respect to the structure of the code, the first unit is the most general, essential one, in which the emission associated to the gaseous component in the simulation is generated without including any additional specification, neither about the observation to be simulated nor about the characteristics of the X--ray instrument, but the theoretical emission model. Ideally, the spectral model assumed can be very complicated and account for more realistic descriptions of the emitting gas which are included in up--to--date sophisticated hydro--simulations, such as metallicity, chemical composition and sub--grid turbulent broadening of the emission lines. 
Conveniently, this first unit can be executed independently from the others just once per simulation output, enormously reducing the computational effort and time. The second unit of the code takes into account the geometry of the problem: the projection along a given line of sight is applied and photon energies are corrected for the Doppler Shift. The order of first and second unit basically reverses the approach of publicly available X--ray virtual telescopes and this has also the main advantage of delaying any limitation on spectral resolution as much as possible, preserving the possibility for any single photon to contribute to the final spectrum.

%
	\subsection{Numerical hydro--simulation input}\label{sec:sims} 
	\name{} requires as input the output of a hydrodynamical simulation.
	For the snapshots of the simulations we use, the redshifts and cosmological parameters are provided as well as the main quantities characterising the gas elements, such as density, temperature, metallicity, position and velocity. 
	Currently, the simulator is best--suited to process simulations performed with SPH, particle--based codes, in particular outputs of the N--body/SPH code GADGET--2 \cite[e.g.][]{springel2001,springel2005}. Nevertheless, the approach is very general and the code can be easily adapted to process also grid--based simulations, for instance performed with Adaptive Mesh Refinement codes like ART \cite[][]{kravtsov1999,kravtsov2002}, ENZO \cite[][and references therein]{norman2007} or FLASH \cite[][]{fryxell2000}.
	In the description of the technique, we will preserve this generality and refer to the gaseous component in terms of more general emitting volumes, namely ``gas elements'', meaning that these could be either particles, as in the application presented in the following Sections \ref{sec:application} and \ref{sec:results}, or grid cells.

%
	\subsection{Unit 1: generation of the box of photons}\label{sec:unit1}
	The first, essential unit is devoted to convert the simulation snapshot from a box of gas elements into a {\it box of photons}. It is crucial to perform this step before the projection is chosen and the emission is integrated along the line of sight in order to achieve and  
	guarantee a very high spatial and spectral precision, accounting for the details of the three--dimensional structure of the source.

	Given the density, temperature and metallicity (or abundances), the spectrum associated with each gas element of the simulation input is generated by means of the publicly available \xspec{} package (v.12)\footnote{See http://heasarc.gsfc.nasa.gov/xanadu/xspec/.}. In particular, since the ICM of galaxy clusters is our primary target, the code assumes either a \mekal{} \cite[e.g.][]{mewe1985,ka_me1993,liedahl1995} or an \apec{} \cite[][]{apec2001} single--temperature, thermal emission model for the hot plasma, with the possibility to add an absorption component given by the \wabs{} model \cite[e.g.][]{wabs1983}. 
	Nonetheless, the code can be readily adapted to account for any desired emission model that is suitable for the considered X--ray source (e.g. \vmekal{}, \vapec{}, \bapec{}, \bvapec{}, for more sophisticated descriptions of the ICM emission including also element abundances or velocity broadening for emission lines). 
	At this stage, we do not convolve the model with any specific instrumental response but we use instead an identity matrix to provide very fine binning over a wide energy band. The choice of the model spectrum adopted, as well as of the energy limits and number of channels for the spectral binning, is left to the user, who can directly specify them in an input parameter file accepted by the code. Additionally, a fiducial angular--diameter distance to the source (${\rm D}_{{\rm A},fid}$) can be specified for convenience as an input parameter, instead of calculating it accordingly to the cosmological redshift of the simulation output.
 
	The main calculation of this first unit consists of computing the emission associated to the gas component processing element per element, which even makes the code naturally suitable to parallelization. 
	Our simulator directly populates the spectrum of each gas element with a distribution of photons, without storing every spectrum or building a library of template spectra. Specifically, we calculate a cumulative spectrum and perform a Monte--Carlo--like simulation to sample it with a total number of photons, ${\rm N}_{ph}$, determined by the total counts expected from the spectral model (in units of ${\rm photons/{\rm s}/\cm^2}$) and the collecting area (${\rm A}_{fid}$) and observing time ($\tau_{{\rm obs},fid}$) specified by the user, so that
	\begin{equation}\label{Nph}
 	{\rm N}_{ph} \propto {\rm A}_{fid} \times \tau_{{\rm obs},fid}.
	\end{equation}
	Ideally, the fiducial values for (${\rm A}_{fid}$) and ($\tau_{{\rm obs},fid}$) have to be assumed in a convenient way such that the spectra are largely over--sampled.

	The photons are therefore characterised by energy (in the emitting element rest--frame), position and velocity. In order to reduce the amount of memory required to store the photon data, photons are organised in terms of packages, each package being associated to an emitting gas element. 
	In this way, photon energies can be stored in a separate output file with respect to position and velocity, which in this first version of the code are equal for all the photons emitted by the same gas element and therefore stored just once per package. 
	Moreover, the data are therefore naturally compressed, since positions and velocities need to be recorded only for gas elements that indeed emit photons. 
	The total number of photons produced from the whole gaseous component in the snapshot processed by \name{} is meant to be largely over--abundant in order to permit a dramatic reduction in the following units because of geometrical selection, projection and instrumental response.

	The data generated from this basic part of \name{} represent a ``cube of photons'' associated to the input hydro--simulation. From this stage on, the simulator has all the information required to proceed with the synthetic observation and the original input is no longer needed. Therefore, it is worth to remark that this first unit can be processed once per output {\it independently} of the specific study to be performed.
%
	\subsection{Unit 2: projection and preparation for the observation}\label{sec:unit2}
	\name{} second unit takes into account the geometry of the observation to be simulated. It accepts the output data produced by the first unit and a parameter file with user--specified parameters for the sub--region of the cube that has to be selected, direction corresponding to the line of sight (l.o.s.), real collecting area and time for the mock observation.

	After selecting geometrically the photons, the energies have to be corrected for the Doppler Shift in order to account for the line--of--sight velocity of the gas element that originally emitted them. In particular, the correction from the emitted to the observed frame is expressed as
	\begin{equation}\label{doppler}
 	E_{ph}^{obs} = E_{ph}^{em}~\sqrt{ \frac{1 - \beta}{1 + \beta} } ,
	\end{equation}
	where $\beta = v_{l.o.s.}/c$ and $v_{l.o.s.}$ is the velocity component along the line of sight. For the sake of simplicity, the line of sight is assumed to be aligned with the $z$--axis. Nevertheless, any different, desired direction only requires a simple rotation of the package positions and velocities before running this second unit of the code.

	The sample of photons obtained has also to be further filtered according to the specified collecting area, ${\rm A},$  and a realistic exposure time, $\tau_{\rm obs},$ chosen to simulate the observation. Accordingly to \eq\ref{Nph}, this is done by calculating the re--scaling factor between these values and the fiducial quantities assumed in the first unit as
	\begin{equation}\label{fak}
 	fak = \frac{{\rm A}~ \tau_{\rm obs}}{{\rm A}_{fid}~ \tau_{{\rm obs},fid}} \times \frac{{\rm D}^2_{{\rm A},fid}}{{\rm D}^2_{\rm A}} ,
	\end{equation}
	and by assuming it as a probability factor for each photon to be actually observed. 
	\eq\ref{fak} explicitly accounts for the possibility of re--sacling the number of observable photons by the angular--diameter distance to the source, ${\rm D}_{\rm A},$ whenever this parameter differs from the fiducial value, ${\rm D}_{{\rm A},fid},$ adopted in Unit 1.

	Lastly, the photon list has to be stored in the most convenient format 
	either to be convolved with the technical response of the specific instrument 
	or to interface with an external software dedicated to simulate specific X--ray satellites.

	This second unit already takes into account the characteristics of the study to be performed but no longer requires the original simulation data, processed by Unit 1. Therefore, any change in the specification of the mock observation can be easily included in this post--processing phase. 
	As such, a straightforward consequence is the unique possibility to investigate the same astrophysical source from many different line of sights, handily processing with Unit 2 the cube of virtual photons obtained with the first unit.
%

	\subsection{Unit 3: simulating the observation}\label{sec:unit3}

	At the third and last stage the mock X--ray observation is completed by eventually considering a real telescope. The photon list obtained from the second unit is convolved with the technical characteristics of a specific instrument, defined by the redistribution matrix file (RMF) and the ancillary response file (ARF). Such process produces final event files which satisfy the standards of real X--ray observations, so that they can be analysed with the same procedures and tools used for real data.

	In the most general perspective this last unit assumes the RMF and ARF files supplied by the user and it is not constructed a priori for a specific instrument. Particular attention, in general, has to be payed to the normalization of the effective area defined by the ARF file of the instrument with respect to the effective area assumed during the projection phase, in order to avoid unphysical overabundance of observed photons.
 
	The third unit of the simulator, being independent of the others, can be conveniently replaced by any desired X--ray instrument simulator, like the sophisticated tool \xissim{}\footnote{See http://heasarc.nasa.gov/docs/suzaku/prop\_tools/xissim/\\xissim\_usage.html.}, which has been developed especially to obtain synthetic observations of the {\it Suzaku} X--ray Imaging Spectrometer \cite[][]{xissim2007}.
%
\section{SYNTHETIC OBSERVATION OF THE COSMIC WEB: two cases of study}\label{sec:application}
We present here an optimal application of \name{} to the hydro--numerical simulation of a filament--like structure. The large--scale region provides indeed a study case to test the remarkable gain in computational cost and time for the post--processing of large data sets; also, the cluster--like haloes residing in this simulated filament offer various science cases to perform the spectral analysis of the mock X--ray observations.
	\subsection{The simulated region}\label{sec:numerics}
	The high--resolution simulated region containing the supercluster--like structure extends for about $50 \times 50 \times 70 \mpc^3$ and consists of 27 haloes at present epoch, four of them being massive cluster--size haloes. This filamentary region has been originally extracted from a cosmological, N--body simulation \cite[][]{jenkins2001,yoshida2001} of a \lcdm universe with $\om = 0.3$, $\omb = 0.044$, $\sigma_8 = 0.9$ and $h = 0.7$, within a box of $479 h^{-1} {\rm (comoving)}~\mpc$ a side. Using the ``zoomed initial condition'' (ZIC) technique \cite[][]{tormen1997}, the overdense region has been re--simulated at higher resolution and a hydrodynamical run with these new initial conditions has been performed with the TreePM/SPH code GADGET--2 \cite[][]{springel2001,springel2005}, including cooling, star formation and feedback from supernova winds. The simulation output contains approximately $1.2 \times 10^{7}$ gas particles and $1.7 \times 10^{7}$ DM particles, at $z=0$, and the final resolution is $m_{DM}=1.30 \times 10^{9}h^{-1}\msun$ and $m_{gas}=1.69 \times 10^{8}h^{-1}\msun,$ for the DM and gas particles, respectively. For an extensive and detailed description of the simulation run and analysis we refer the reader to the paper by \cite{dolag2006}.
	\subsection{Suzaku mock observations with \name{}}\label{sec:main}
	We applied \name{} to the whole simulation output corresponding to redshift $z = 0.07$ and obtained from the first unit a list of all the photons emitted by the sufficiently--hot gas particles in the simulated region. In order to compute the model spectra, we adopted an \apec{} model with fixed metallicity, set for simplicity to $Z = 0.3\zsun$ for all the particles, and assumed the solar abundances by \cite{angr1989}. 
	The value assumed for the Galactic equivalent column density for the \wabs{} absorption model was $N_H = 7 \times 10^{20} \cm^{-2}.$ 
	The values for fiducial collecting area and exposure time were initially set to $2000\cm^2$ and $1 {\rm Ms}$, respectively. This first stage was the most computationally demanding (see \tab{}\ref{table:timing} for time scales and memory requirements referring to the application presented here) and permitted to generate approximately $1.8 \times 10^8$ photons, for the whole simulation output.
	\begin{table}
          \begin{center}
            \caption{Typical computational time scales and output storage memory for the hydrodynamical numerical simulation and different units of \name{}. As a representative case, the data reported here refer to the science application$^{\dag}$ presented in \sec\ref{sec:application}.}
            \begin{tabular}{cccc}
              \hline
              {\bf Sim} &  & {\bf PHOX} &  \\
              \hline
              & {\bf Unit 1} & {\bf Unit 2} & {\bf Unit 3/}\xissim{} \\
              \hline
              \hline
	& $\sim$1h\,40min$^*$ & $\sim$few sec & $\sim$few min\\
	\hline
	1.8~Gb & 827~Mb & 176~Mb (\leiaB{}) & $\sim$100~Mb (\leiaB{}) \\
        &        & 60~Mb (\leiaD{}) & $\sim$35~Mb (\leiaD{})\\
	\hline
      \end{tabular}
    \end{center}
    \label{table:timing}
    \scriptsize{$^*$The run reported here has been performed by binning in temperature the particles (assuming $\Delta{\rm (kT)} \sim 5 eV$) and calculating a spectrum per bin, rescaling afterwards the expected total number of photons by the specific normalization calculated for each particle in the temperature bin. Since no varying metallicities have been considered in this run, this approach has been used to further reduce the computational effort.\\ \\{$^{\dag}$The results are obtained for serial runs performed on standard work station (2300 MHz, AMD Opteron).} }
  \end{table}
	\begin{figure*}
	\begin{center}
	\includegraphics[width=0.67\textwidth]{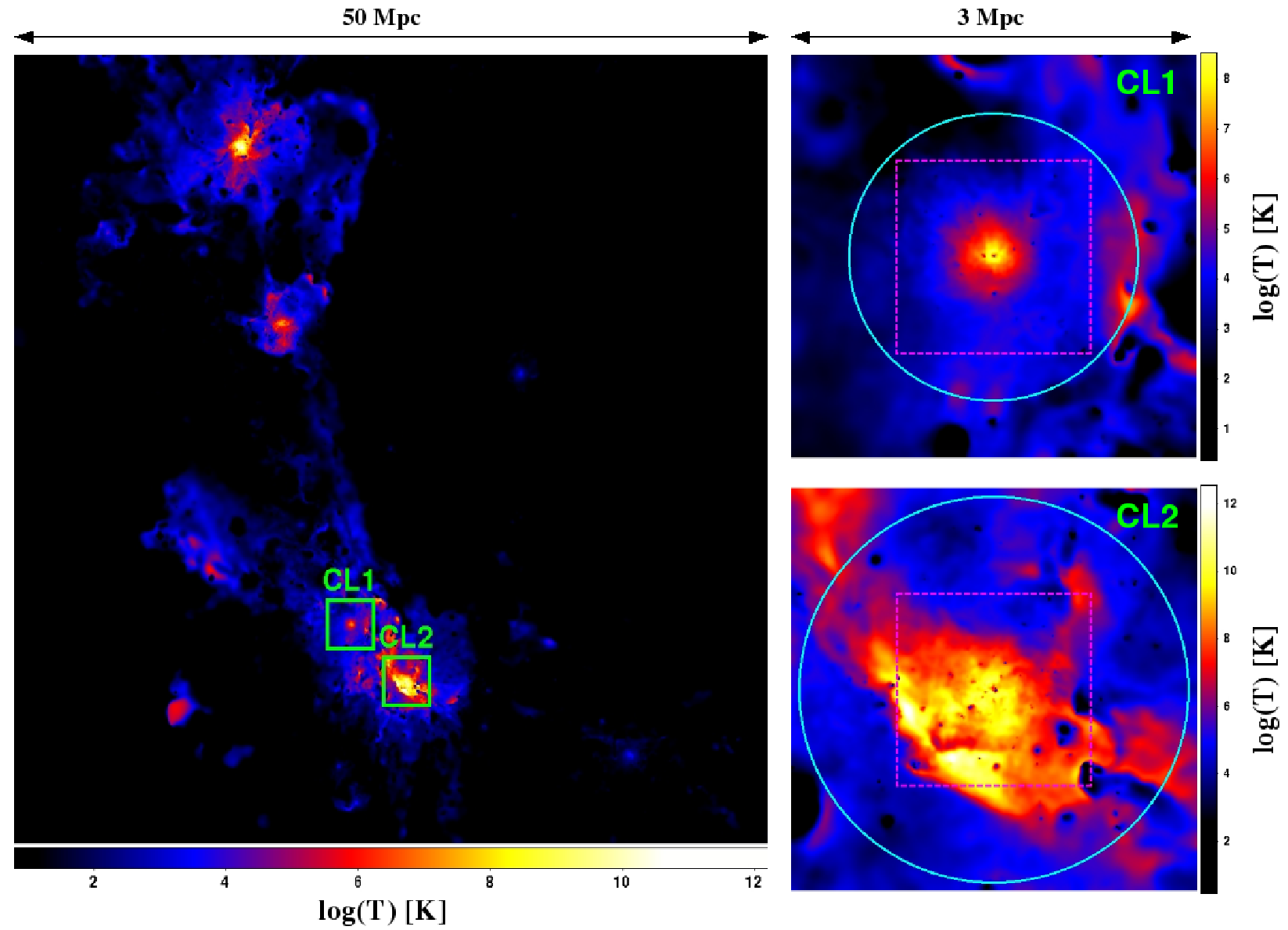}
	\includegraphics[width=0.32\textwidth]{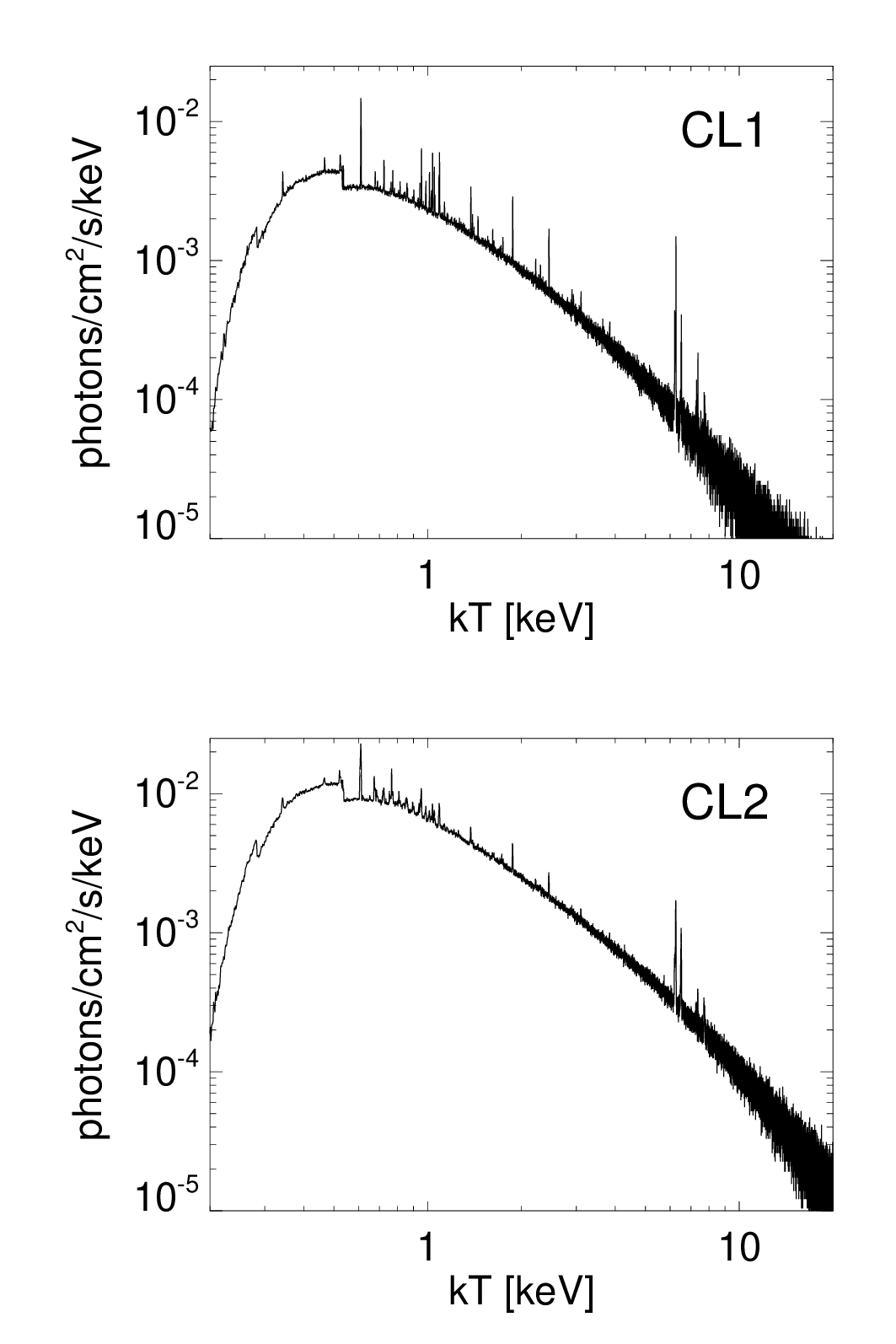}
	\caption{Filament--like region and zoom onto the haloes \leiaD{} and \leiaB{}, at $z=0.07$. Left--hand panel: projection along the $z$--axis of the emission--weighted temperature, in a logarithmic color scale. The map is $50 \mpc$ a side and encloses the high--resolution region of the box containing the filament. The $3 \times 3 \mpc^2$ regions centered on \leiaD{} and \leiaB{}, respectively, are marked with the green squares. Central column: zoom onto \leiaD{} (top panel) and \leiaB{} (bottom panel). The maps are $3 \mpc$ a side and show the logarithm of the emission--weighted temperature, projected along the $z$-axis. The $\rfive$ region (cyan circle) and the XIS FOV (magenta dashed square) are overplotted for both haloes. Right--hand column: theoretical spectra obtained from Unit 1 and 2 of \name{} for the $\rfive$ region of both haloes.} 
	\label{fig:leiaSB}
	\end{center}
	\end{figure*}

	In Unit 2, we assumed to observe the large--scale region from a physical distance\footnote{We set the observer on the positive part of the $z$--axis, assumed to be aligned with the l.o.s..} of $313.9 \mpc$, which corresponds to the luminosity distance at the snapshot redshift $z=0.07.$ Similarly, with the given cosmological parameters, the angular--diameter distance at this redshift is $274.4 \mpc,$ which gives a physical scale of $1.33 \kpc/{\rm arcsec}$. In this projection phase, we select two specific sub--regions of the simulation box, containing two cluster--like haloes among the four most massive ones. In particular, we consider for the present analysis a regular, massive cluster ({\it hereafter}, \leiaD{}) with $\mfive=3.97 \times 10^{14}\msun$ and $\rfive=1070.5 \kpc$  and a massive, disturbed system with $\mfive=9.24 \times 10^{14}\msun$ and $\rfive=1417.3 \kpc$ ({\it hereafter}, \leiaB{}). The region geometrically selected around both haloes is a cylinder along the $z$--axis, enclosing $\rfive$ in the $xy$ plane. Planning to simulate a Suzaku observation, the collecting area assumed for the mock observation equals the physical area of the {\it Suzaku} X--ray Imaging Spectrometer (XIS), ${\rm A} = 1152.41 \cm^2$, and the exposure time is $\tau_{obs} = 500 {\rm ks}$, in order to have good statistics. This allows to extract $3.1 \times 10^6$ photons for \leiaD{} and $9.2 \times 10^6$ photons for \leiaB{}.

	In the left--hand panel of \fig\ref{fig:leiaSB} we show the filament--like structure at $z=0.07$, projected along the $z$--axis (i.e. the observation l.o.s.). The map represents the emission--weighted temperature of the simulated filament enclosed in the high--resolution region. Overplotted in green, we mark the zoom onto the selected haloes, which are shown in the central column of \fig\ref{fig:leiaSB}. The two panels contain in fact the $3 \times 3 \mpc^2$ maps of the emission--weighted temperature projected along the l.o.s., for \leiaD{} and \leiaB{}. For comparison, we overplot also the $\rfive$ region (cyan circle) and the XIS field of view (magenta dashed square). Instead, in the right--hand column of \fig\ref{fig:leiaSB} we show the ideal integrated spectra for \leiaD{} and \leiaB{}, as obtained from Unit 1 and 2 after selecting the photons coming from the region enclosed within $\rfive$. These spectra are obtained purely through a regular binning of the photon energies with bins of $\Delta E = 0.001 \kev$, without including any sensitivity or effective area.

 	For the purpose of this paper, we present \name{} by taking advantage of the public package \xissim{} \cite[][]{xissim2007}, designed to simulate observations with the XIS spectrometer on board the {\it Suzaku} X--ray satellite. By adapting the second unit of our code to produce the output photon list in a format best--suited for \xissim{}, we test our novel technique and demonstrate its capabilities through a simple science application.

	Specifically, we use \xissim{} to process the simulated photons generated by \name{} with the real responses and calibration characteristics of different XIS CCDs. Given the high statistics offered by our simulation, we decide to consider the back--illuminated (BI: XIS-1) sensor and the two combined, front--illuminated CCDs (FI: XIS-0,\,-3) separately. The combination of two $500 {\rm ks}$ observations with the XIS-0 and XIS-3 detectors additionally improves the statistics for the final FI spectrum.

	The event files obtained for the two cluster--like haloes of the filament, \leiaD{} and \leiaB{}, without any addition of physical background emission, were then analysed following the standards of X--ray data analysis.
	%
	\subsection{Spectral Analysis}\label{sec:xspec}
\begin{figure*}
	\centering
	\includegraphics[width=0.32\textwidth]{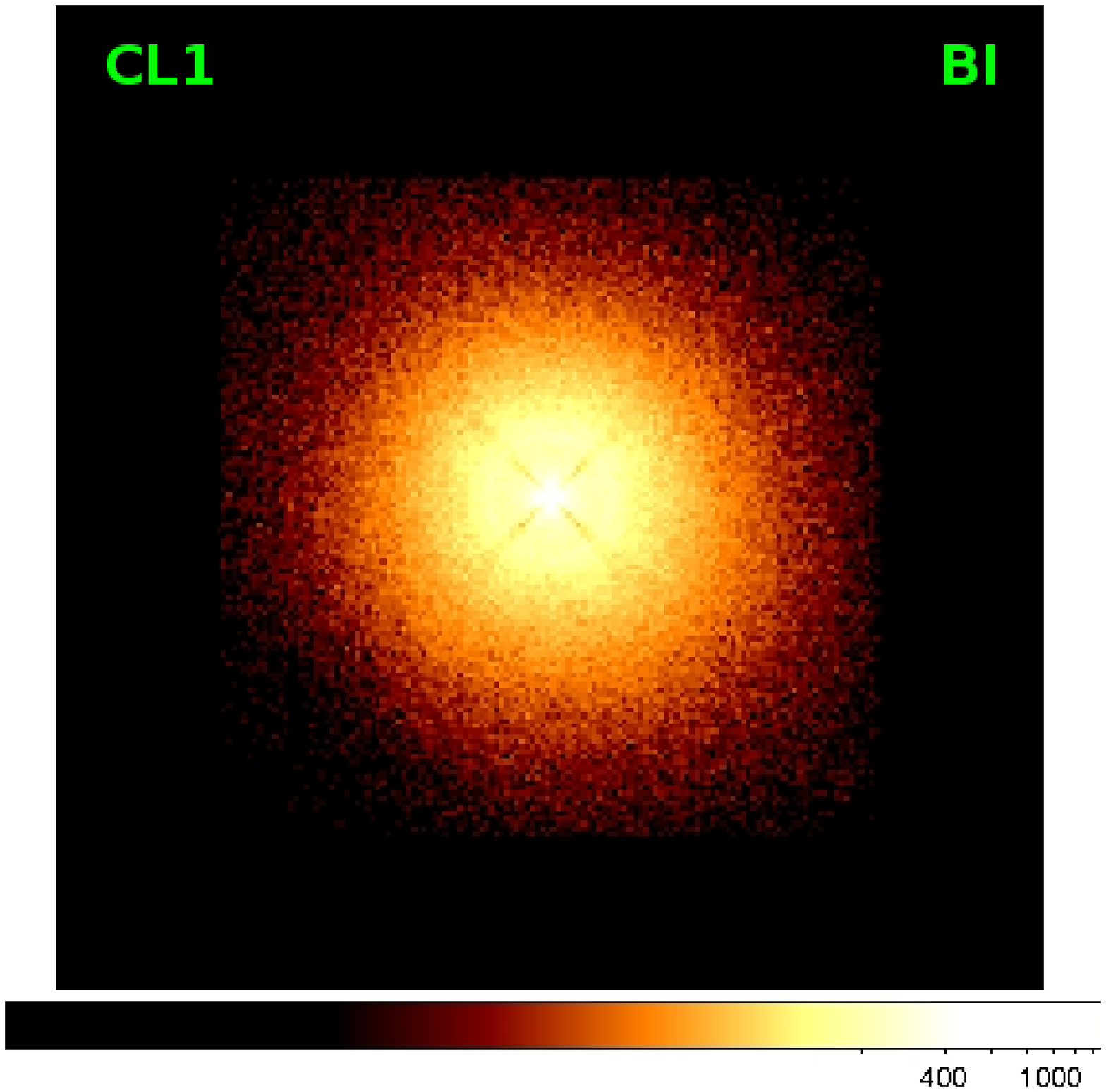}
	\includegraphics[width=0.41\textwidth]{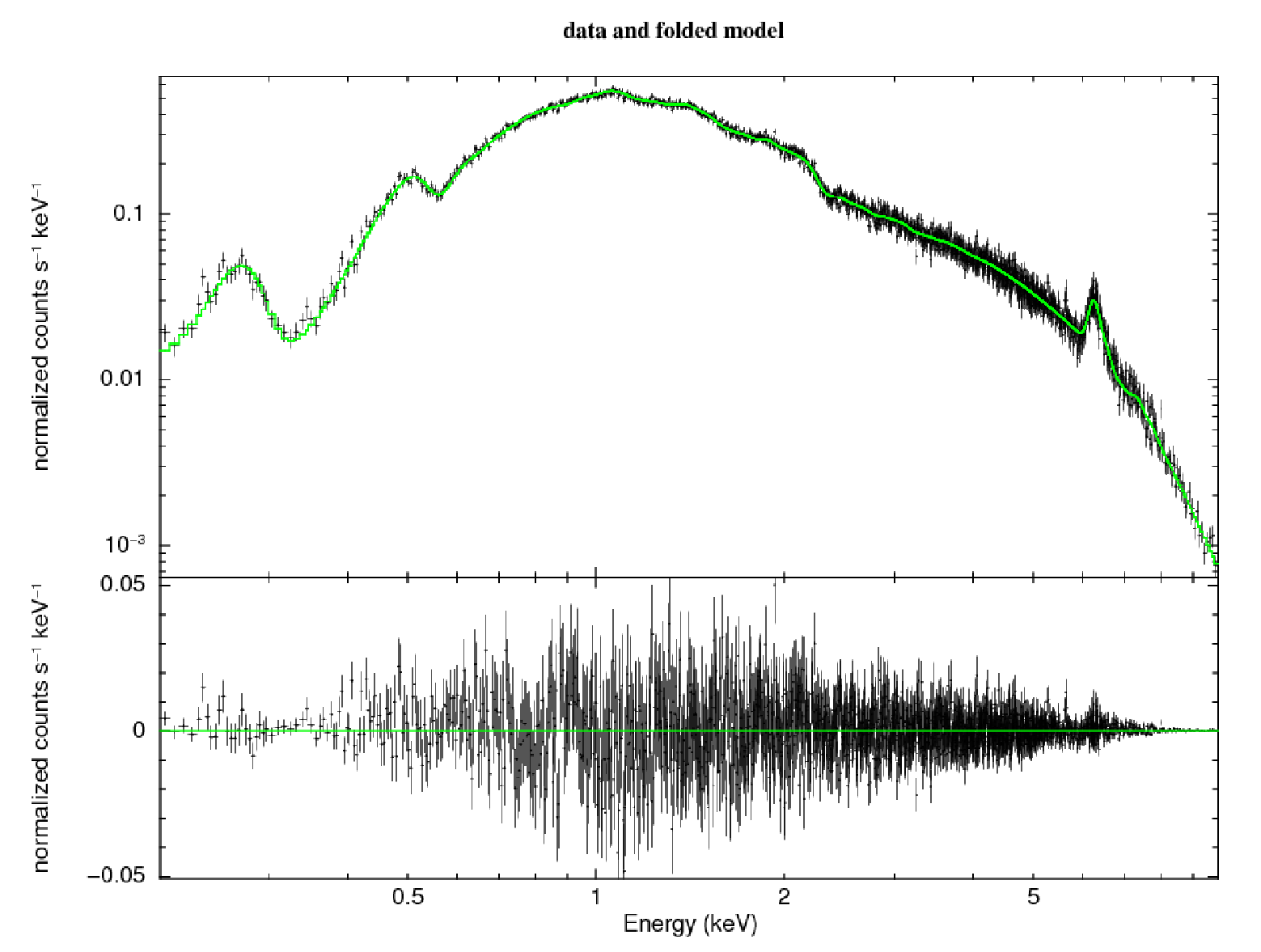}
	\includegraphics[width=0.32\textwidth]{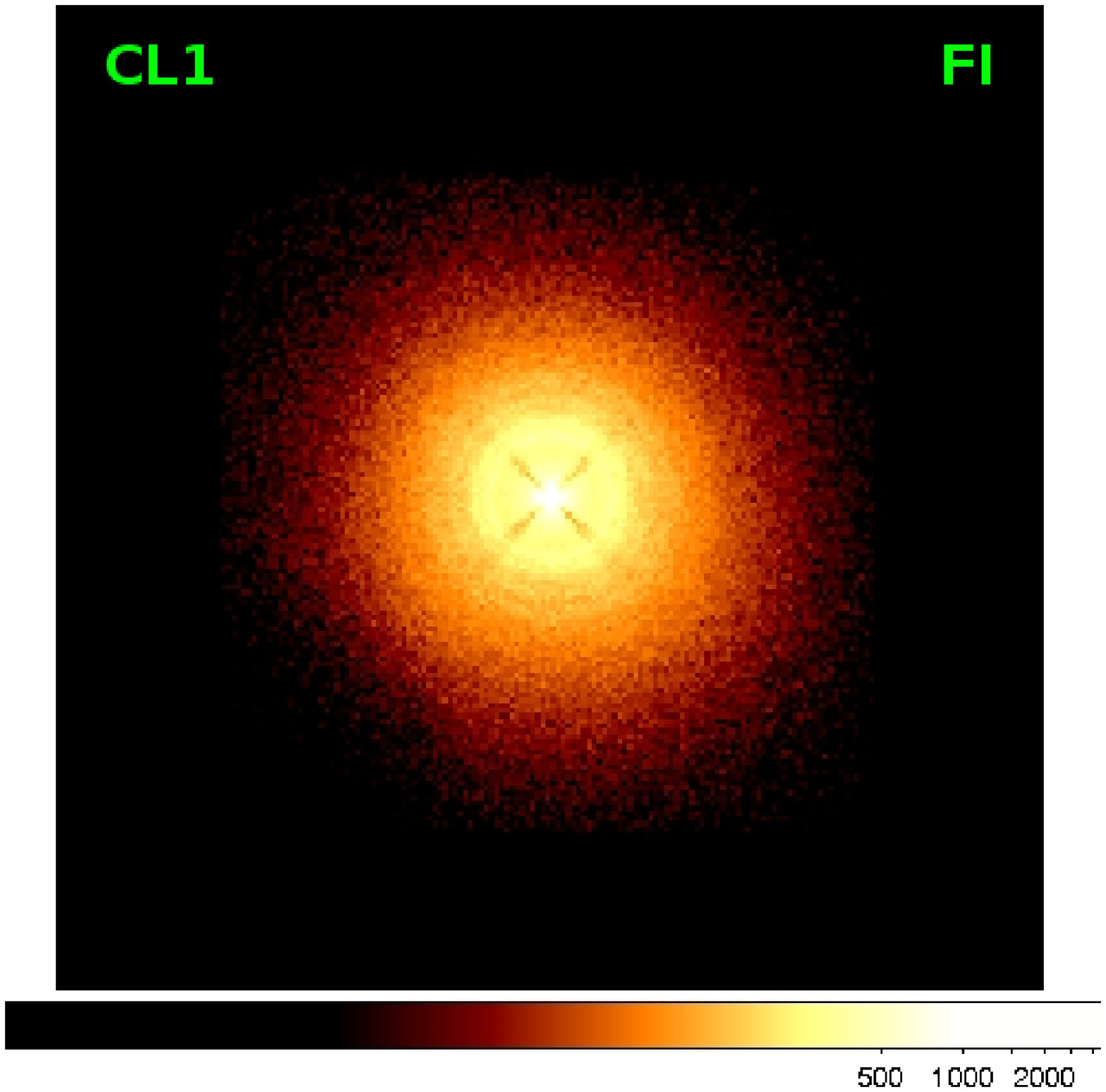}
	\includegraphics[width=0.41\textwidth]{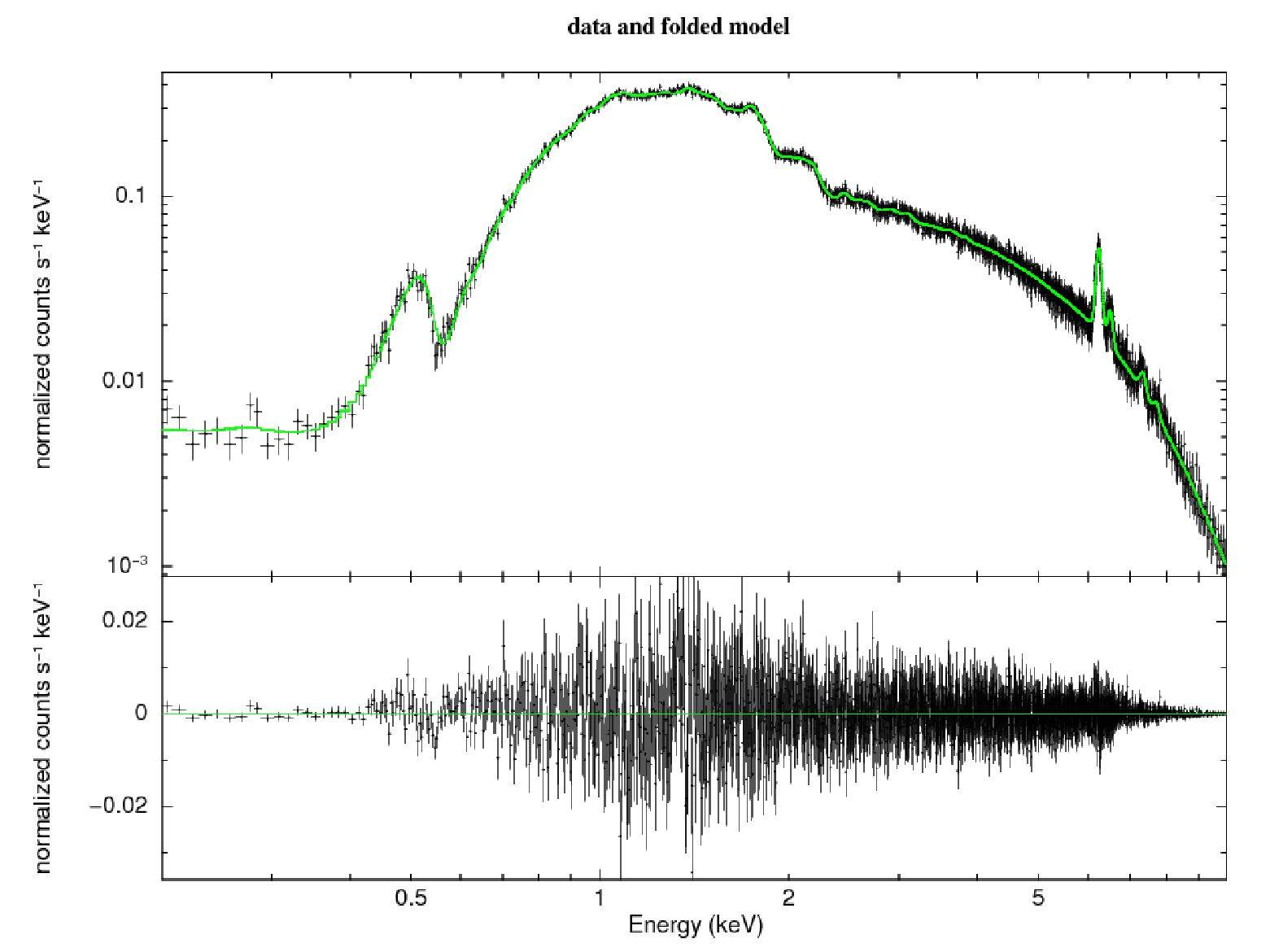}
	\includegraphics[width=0.32\textwidth]{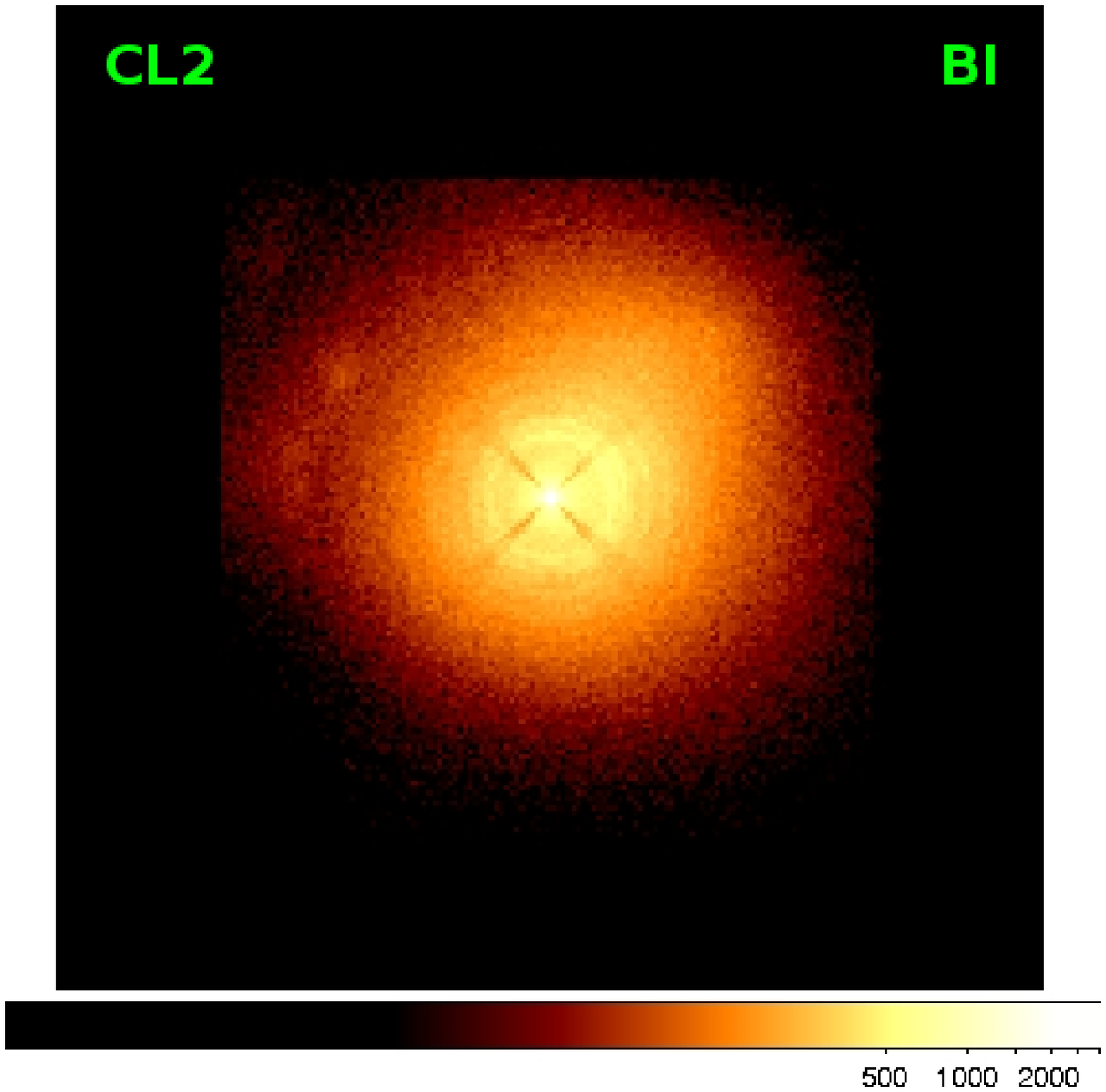}
	\includegraphics[width=0.41\textwidth]{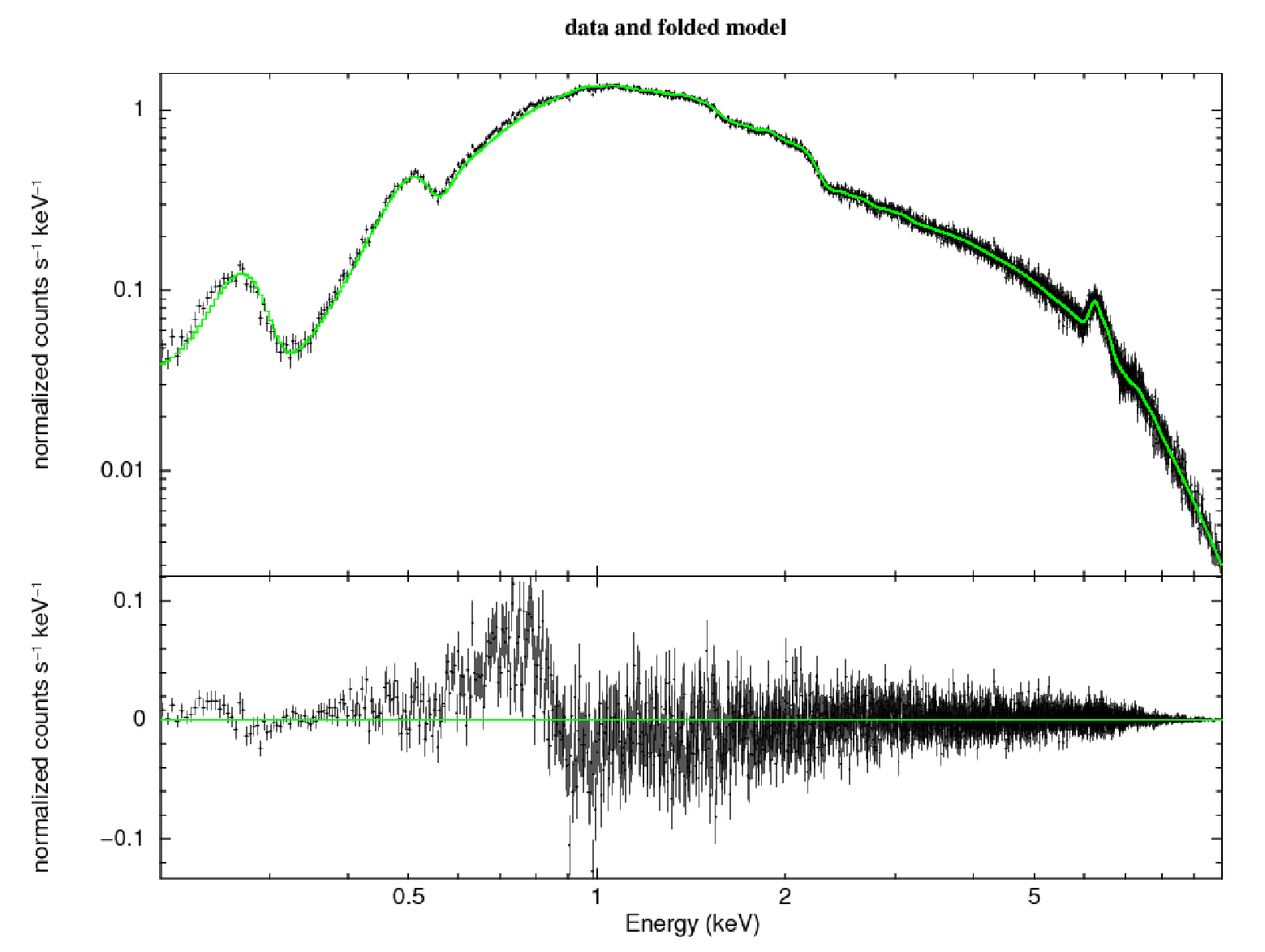}
	\includegraphics[width=0.32\textwidth]{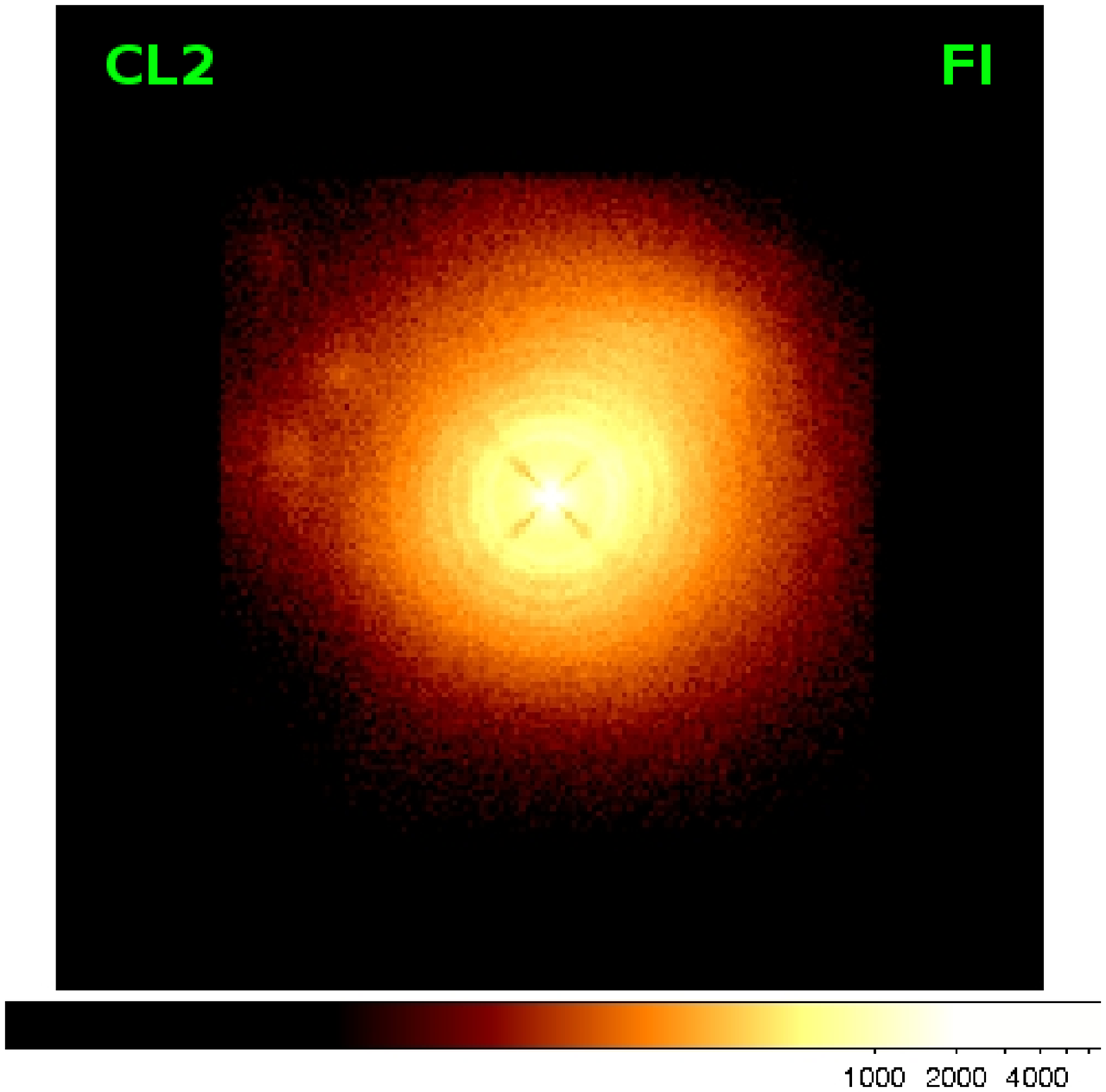}
	\includegraphics[width=0.41\textwidth]{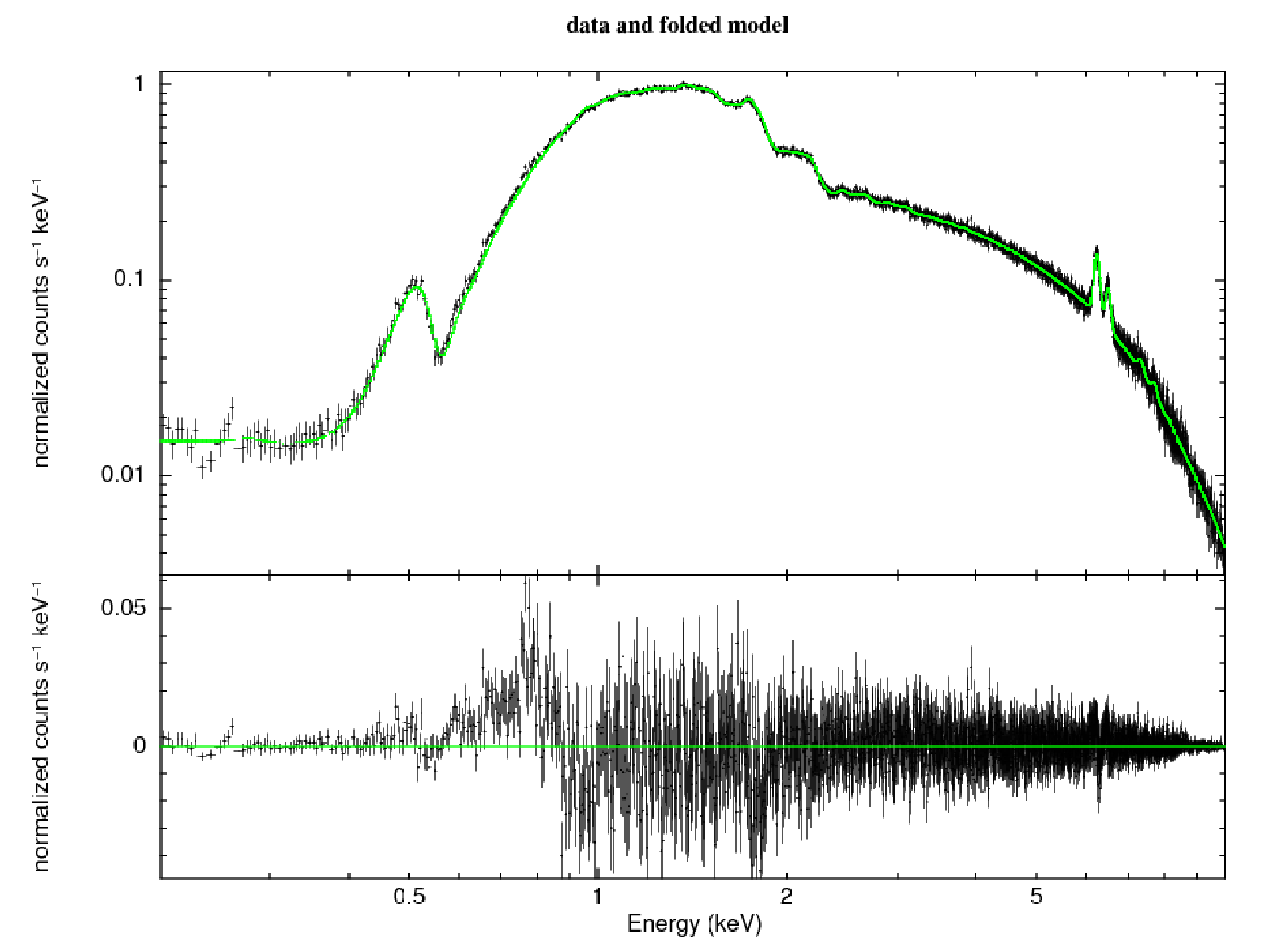}
	\caption{Simulated {\it Suzaku} images (left--hand column) of \leiaD{} and \leiaB{} and corresponding spectra (right--hand column). Also the models fitted to the spectra are shown. For each halo the BI XIS and the combined FI XIS sensors are presented.}
	\label{fig:xisimg}
	\end{figure*}

	In \fig\ref{fig:xisimg} we show the simulated photon images (left--hand column) of the $500{\rm ks}$ XIS observations, with the BI and FI sensor, for \leiaD{} and \leiaB{}. The images, as well as the spectra, were extracted from the original event files using {\tt Xselect} (v.2.4) from the FTOOLS\footnote{ See http://heasarc.gsfc.nasa.gov/ftools/.} package \cite[][]{ftools}, and correspond to the central region in both clusters. In particular, the {\it Suzaku} XIS covers a field of view (FOV) of roughly $18' \times 18'$, which corresponds to a physical scale of $1436.6\kpc$ at the distance of our sources and encloses therefore the region within $\sim 0.67\rfive$ of \leiaD{} and $\sim 0.51\rfive$ of \leiaB{} (we refer to the central panels of \fig\ref{fig:leiaSB} for a visual representation of the $\rfive$ region and the XIS FOV in the two haloes).

%

	\begin{table*}
	\begin{center}
	\caption[Xspecfit]{Multi--temperature best--fit results for {\it Suzaku} XIS simulated spectra of haloes \leiaD{} and \leiaB{}. The model adopted is $\wabs{} \times (\apec{}_1 + \apec{}_2 + \apec{}_3 + \apec{}_4 + \apec{}_5),$ where only the normalizations $K_i$ are free in the fit.}
	\begin{tabular}{cccccccc}
	\hline
	XIS & $K^{\star}_1$ & $K_2$ & $K_3$ & $K_4$ & $K_5$ & $\chi^2/d.o.f.$\\
	\hline
	\hline
	\multicolumn{7}{c}{ \leiaD{} }\\
	\hline
 	& $(kT_1=0.69\kev)$ & $(kT_2=1.38\kev)$ & $(kT_3=2.75\kev)$ & $(kT_4=5.5\kev)$ & $(kT_5=11\kev)$ & \\
	\hline
	BI & $0.21(\pm0.04)\times 10^{-3}$ & $0.15(\pm0.16)\times 10^{-3}$ & $2.03(\pm0.60)\times 10^{-3} $ & $8.17(\pm0.63)\times 10^{-3} $ & $1.38(\pm0.54)\times 10^{-3} $ & $1693.2/1700$\\
	FI & $0.27(\pm0.04)\times 10^{-3}$ & $0.12(\pm0.14)\times 10^{-3}$ & $2.00(\pm0.44)\times 10^{-3} $ & $8.12(\pm0.43)\times 10^{-3} $ & $1.47(\pm0.36)\times 10^{-3} $ & $1713.2/1998$\\
	\hline
	\hline
	\multicolumn{7}{c}{ \leiaB{} }\\
	\hline
 	& $(kT_1=1\kev)$ & $(kT_2=2\kev)$ & $(kT_3=4\kev)$ & $(kT_4=8\kev)$ & $(kT_5=16\kev)$ & \\
	\hline
	BI & $0.99(\pm0.09)\times 10^{-3}$ & $0.0(\pm0.91)\times 10^{-3}$ & $7.87(\pm1.70)\times 10^{-3} $ & $16.44(\pm2.28)\times 10^{-3} $ & $7.87(\pm1.41)\times 10^{-3} $ & $2762.0/2167$\\
	FI & $1.16(\pm0.09)\times 10^{-3}$ & $0.0(\pm0.70)\times 10^{-3}$ & $7.49(\pm1.21)\times 10^{-3} $ & $16.14(\pm1.52)\times 10^{-3} $ & $8.32(\pm0.90)\times 10^{-3} $ & $2557.3/2435$\\
	\hline
	\end{tabular}
	\end{center}
	\begin{flushleft}
	$^{\star}$The normalization of an \apec{} component is defined as $K = 10^{-14} \int n_e n_H dV / [4\pi D^2_A (1+z)^2]~\cm^{-5},$ where $D_A$ is the angular--diameter distance to the source.
	\end{flushleft}
	\label{table:xspecfit}
	\end{table*}
	Simulations of cluster--like haloes provide the possibility of precisely knowing the intrinsic dynamical and thermal properties of the ICM. From a visual inspection of the temperature maps shown in \fig\ref{fig:leiaSB} (central column), the different structure of the two haloes studied is already evident. As we will discuss in more detail in \sec\ref{sec:results} (see \fig\ref{fig:leiaDBem}), the further investigation of the emission measure ({\it hereafter}, E.M.) distribution as function of temperature, for the gas particles in the simulation that reside in the FOV of {\it Suzaku} both in \leiaD{} and \leiaB{}, let us unveil the halo intrinsic thermal structures. Especially, \leiaD{} is mainly dominated by one temperature component, although the E.M. distribution is not very narrow, whereas the second halo, \leiaB{}, clearly has a complex thermal structure, which cannot be well described by a single temperature.

	Therefore, we fit a multi--temperature model to the BI and FI XIS spectra, restricting our analysis to the $0.2-10~\kev$ energy band and requiring a minimum of $50$\footnote{The spectra were rebinned using the {\tt grppha} routine from the FTOOLS package.} counts per energy bin. The spectral fit was performed with \xspec{} 12.6.0 \cite[][]{xspec1996}. The response files (RMFs) used in the fit were the same as in the run of \xissim{}, while we generated the ARF files for each halo and detector by means of the ftool \arfgen{} \cite[][]{xissim2007}.

	For our spectral modelling of the simulated data, we considered a $\wabs{} \times (\apec{}_1 + \apec{}_2 + \apec{}_3 + \apec{}_4 + \apec{}_5)$ model, describing each component by an \apec{} model, as assumed in Unit 1 for the emission associated to each gas element. The equivalent hydrogen column density for the Galactic absorption model \wabs{} was frozen to $N_H = 7 \times 10^{20} \cm^{-2},$ as in the original run of Unit 1. For analogous reasons, we adopted fixed values for redshift, $z=0.07,$ and metallicity $Z = 0.3\zsun,$ always assuming the \cite{angr1989} solar abundances.

	In both haloes, we assigned to the first temperature component, $T_{\ast},$ a value fairly close to the emission--weighted temperature estimated from the gas particles in the simulation. Then, following a strategy similar to what is suggested by \cite{peterson2003} or \cite{kaastra2004}, we assigned to the other components temperatures of $2T_{\ast}$ and $\frac{1}{2}T_{\ast},\frac{1}{4}T_{\ast},\frac{1}{8}T_{\ast},$ for the cooler ones. The five normalizations of the \apec{} components where free parameters in the fit.

	The spectra and best--fit models are shown in the right--hand column of \fig\ref{fig:xisimg}, for the two haloes and the two XIS sensor kinds. The results of the spectral fit are presented in \tab\ref{table:xspecfit}. The five temperatures, frozen in the fit, are different for \leiaD{} and \leiaB{} and therefore reported in the table for clarity reasons. Overall, the resulting $\chi^2_{red}$ is always very good, meaning that the five--temperature model provides a fair description of the thermal structure for both clusters in the central observed region.
	%
\section{Results: recovering the emission measure distribution}\label{sec:results}
\begin{figure*}
\includegraphics[scale=0.49]{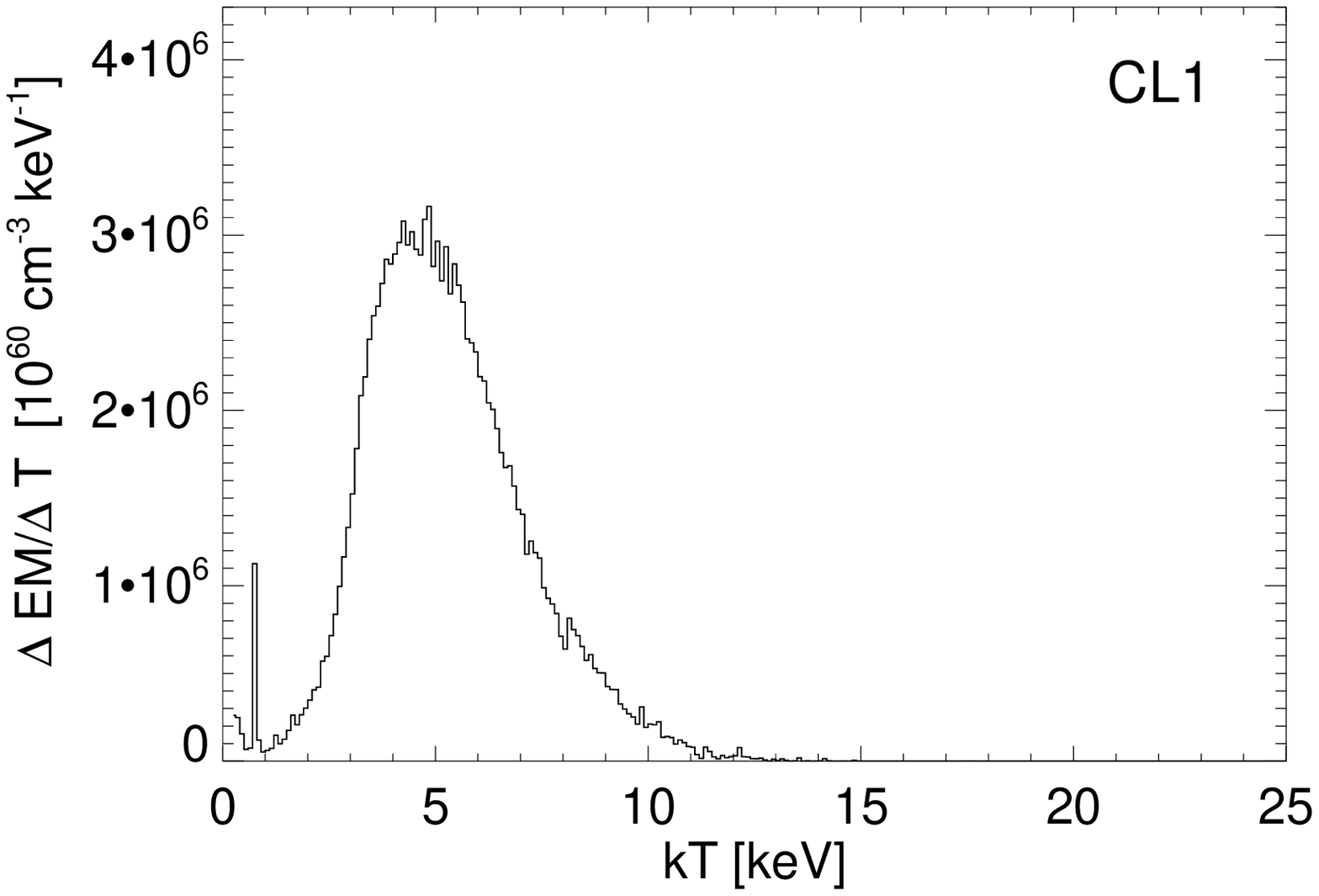}
\includegraphics[scale=0.49]{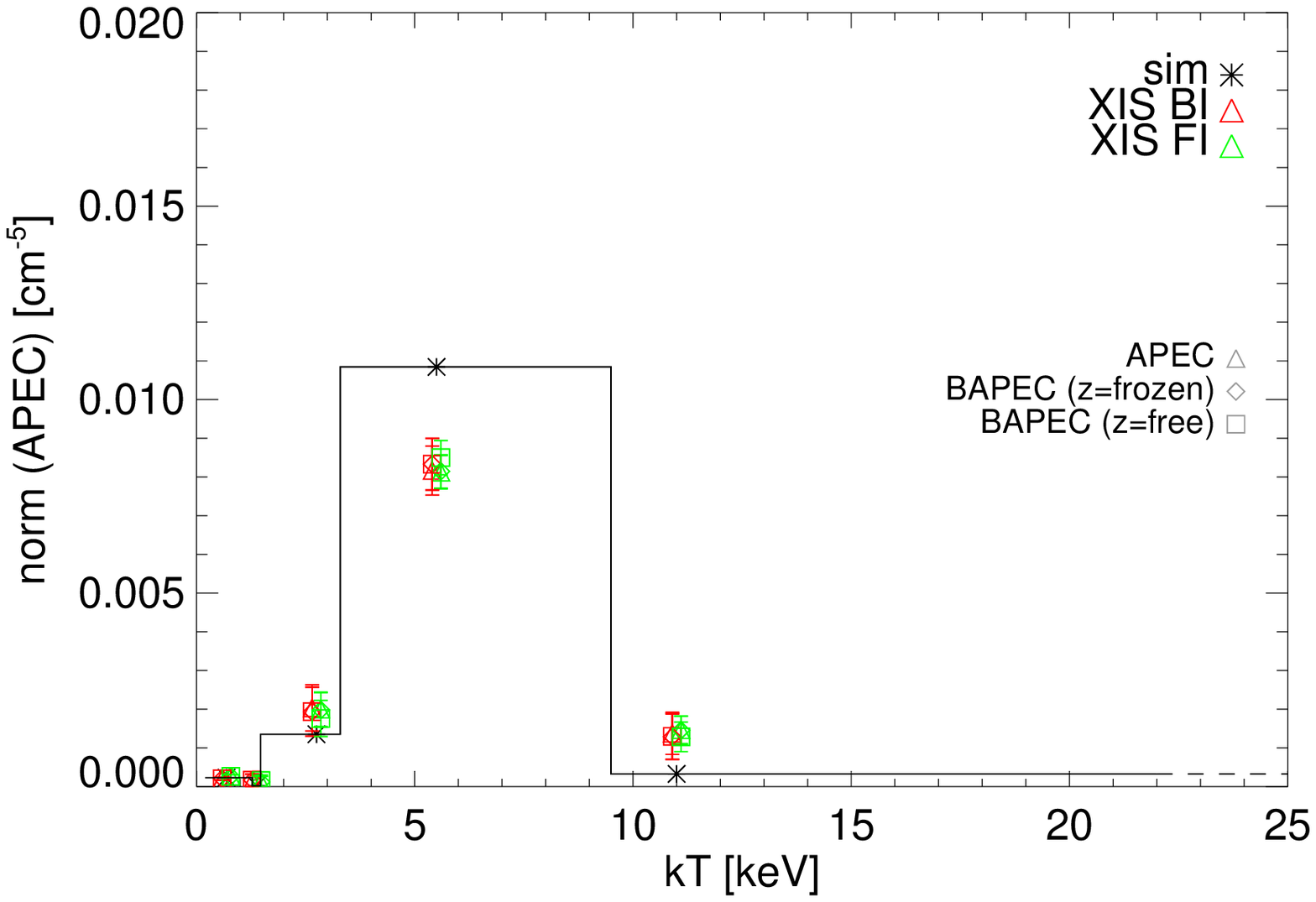}
\includegraphics[scale=0.49]{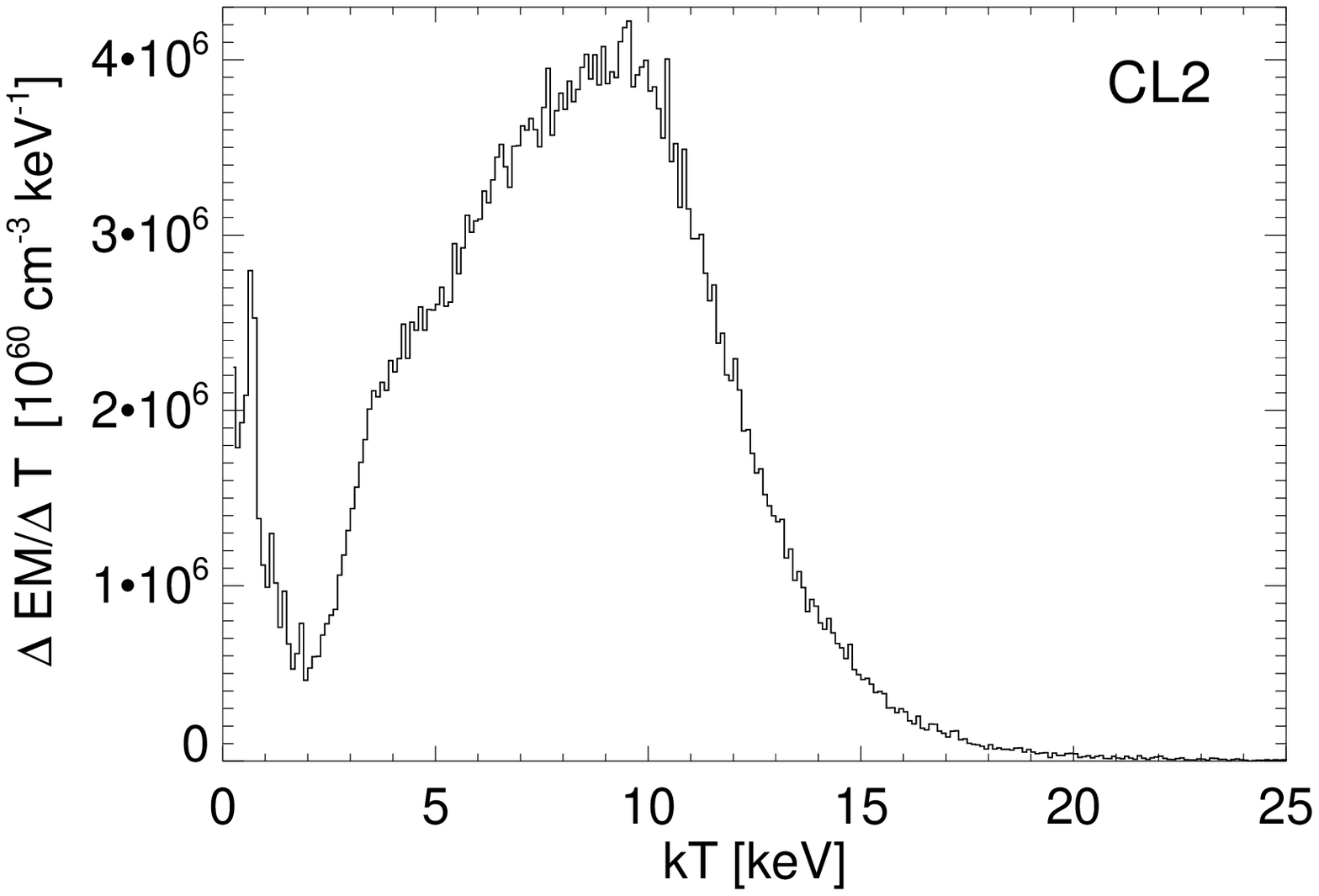}
\includegraphics[scale=0.49]{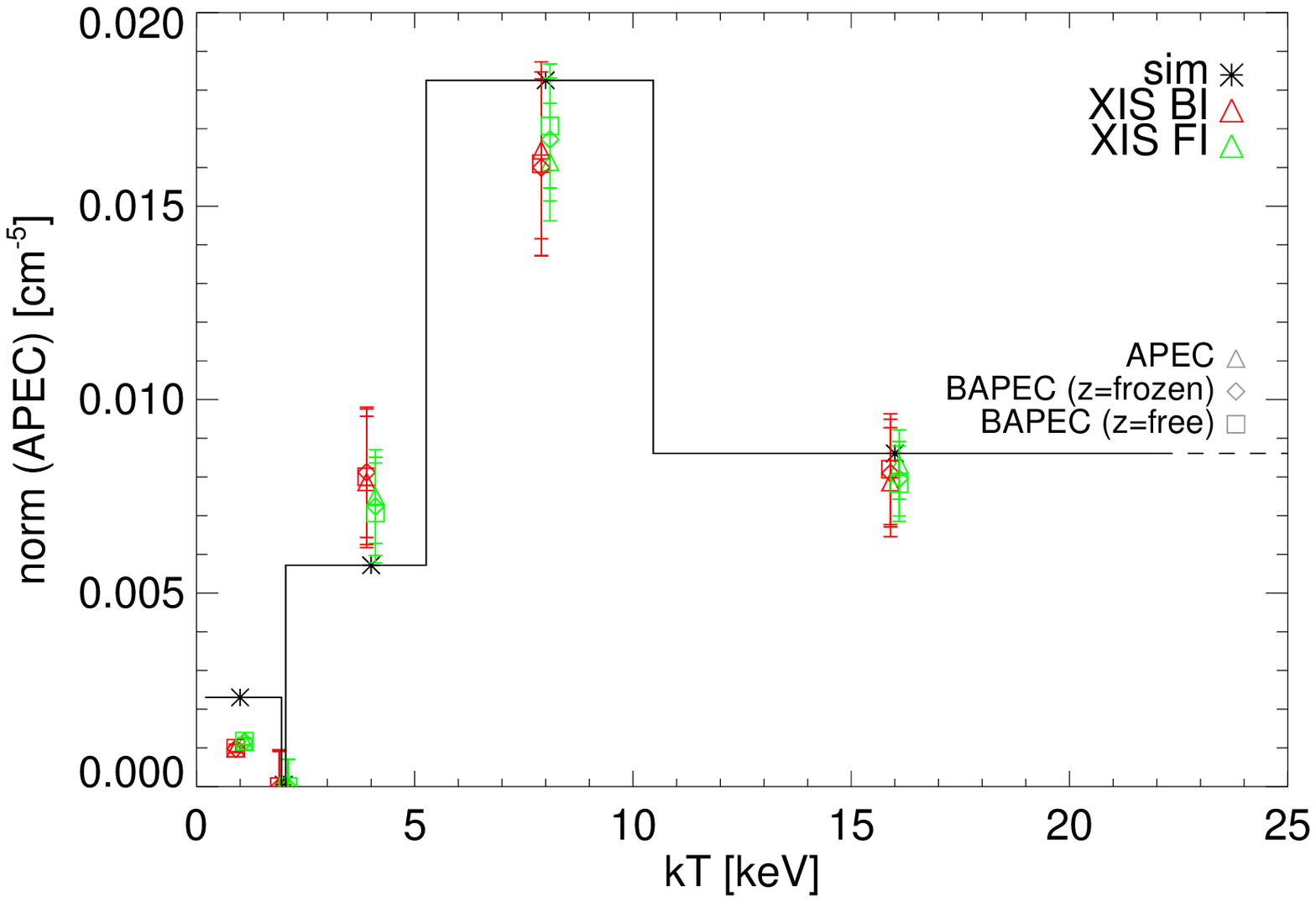}
\caption{Left--hand column: emission measure (E.M.) distribution as a function of temperature for gas particles contained in the FOV of Suzaku XIS (projected along the $z$--axis), for haloes \leiaD{} and \leiaB{} at redshift $z=0.07$. Right--hand column: comparison between best--fit values of APEC (triangles) and BAPEC (--1, diamonds and --2, squares) normalizations and corresponding theoretical values (asterisks), computed from the E.M. distribution of emitting gas particles. Different colors for the best--fit values refer to spectra obtained with different XIS detectors: BI XIS (red), and FI XIS (sum of XIS-0 and XIS-3, green). In both columns, the upper row refers to \leiaD{} and the lower row refers to \leiaB{}.}
\label{fig:leiaDBem}
\end{figure*}
The interesting purpose of our multi--temperature modelling is the reconstruction of the thermal structure of the emitting ICM. This is pursued by directly comparing the distribution of the best--fit normalizations at the corresponding, fixed temperatures with the expected values calculated from the simulation. Dealing with simulated data, we can faithfully test the results of the spectral analysis against the underlying solution in a straightforward way, which is not possible for real data.

By definition, the normalization of each \apec{} component is
\begin{equation}\label{norm}
	K = 10^{-14} {\rm E.M.} / [4\pi D^2_A (1+z)^2]~\cm^{-5},
\end{equation}
where $D_A$ is the angular--diameter distance to the source and $z$ is the redshift. This implies that the normalization is directly proportional to the E.M. of the hot emitting plasma, defined as
\begin{equation}\label{em}
	{\rm E.M.} = \int n_e n_H dV,
\end{equation}
where $n_e$ and $n_H$ are the electron and hydrogen number densities of the plasma, respectively.
In order to map the distribution of E.M. in the simulated haloes to the corresponding $K$ values at fixed temperatures, the theoretical temperature binning has to be determined in the most faithful way possible.

The shape of the E.M. distribution extracted from the simulation is presented for the two haloes in the left--hand panels of \fig\ref{fig:leiaDBem}. This has been calculated considering all the gas particles of the simulated haloes, for which the projected position on the $xy$ plane is enclosed in the {\it Suzaku} field of view. A lower temperature cut has been artificially set to $T_{min} = 0.2 \kev,$ which is the lower boundary of the energy band used in the spectral fit. Nevertheless, this continuous distribution has to be reduced to the five fixed temperatures adopted for the spectral fit. The most consistent and independent way of choosing the boundaries between the different temperature components, is by requiring that the value of the 
E.M.--weighted temperature for each bin equals the fixed value adopted for the fitting (we follow here an approach analogous to the one adopted by \cite{kaastra2004}). 
Specifically, we iteratively adjust the division boundaries until the match between the E.M.-weigthed temperature in the bin and the fixed value is reached, starting from the lower--temperature component and proceeding towards the higher ones. 
Finally, the total E.M. calculated in each bin can be converted into an expected value for the \apec{} normalization via \eq\ref{norm}.

The direct comparison between simulation and mock--observation results is presented in the right--hand panels of \fig\ref{fig:leiaDBem}, for both \leiaD{} and \leiaB{}. The original E.M. distribution for the two haloes is plotted in the left--hand panels of \fig\ref{fig:leiaDBem}, where it is clear that none of them is strictly isothermal, although the \leiaD{} cluster has a narrower thermal distribution with respect to \leiaB{}. The theoretical expectations for the fixed temperature components are plotted as black asterisks on the right--hand panels, where the best--fit values are reported as well.

Interestingly, the agreement between simulated and ``observed'' values is quite good and the overall shape of the thermal distribution in both clusters is recovered reasonably well. This result is found to be true for both the BI and FI XIS sensors. 
Moreover, the differences between the thermal structures of the two haloes are reflected by a different width and extension of the E.M. distribution over temperature, which is definitely broader in the \leiaB{} case.

We notice that the dominant component at $kT_4=5.5\kev$ is underestimated by the fit for the halo \leiaD{}. 
Nonetheless, it is worth to notice that the determination of the most consistent temperature binning of the theoretical E.M. distribution is expected to be sensitive to the definition of temperature assumed to match the temperature of the thermal component in the spectral fit. For instance, it is known that there is a discrepancy between the emission--weighted temperature, as usually calculated in the analysis of numerical simulations, and the projected spectroscopic temperature obtained from X--ray observations \cite[see, for instance,][]{mazzotta2004}. In this respect, our choice for the theoretical temperature calculated from the simulation, which is weighted by the E.M., is believed to be the most consistent. 
However, we also tested the results against a different definition of temperature, namely the emission--weighted\footnote{The emission--weighted temperature is defined as $$T_{ew} = \frac{\int n^2 \Lambda(T) T dV}{\int n^2 \Lambda(T) dV}, $$ where $n$ is the gas density and $\Lambda(T)$ is the cooling function. As usually done, we consider here $\Lambda(T) \propto T^{\frac{1}{2}},$ assuming the gas to emit mainly via Bremsstrahlung.} one, which resulted in a slightly different set of theoretical values for the five \apec{} normalizations. Even in this case, our conclusions are not affected in any significant way and the overall shape of the underlying E.M. distribution is equally well traced by the multi--component spectral fit.

In the right--hand plots of \fig\ref{fig:leiaDBem} we also show two alternative five--temperature models adopted to fit the synthetic spectra. Specifically, we perform two spectral fits with a multi--temperature model consisting of five \bapec{} components. The \bapec{} spectral model is a velocity-- and thermally--broadened emission spectrum, where the velocity broadening is parametrized by a gaussian velocity dispersion $\sigma_v$. In fact, the technique implemented in our simulator allows to preserve the very high energy resolution of the integrated spectrum till the final convolution with the instrument. Since every photon energy carries the imprint of the emitting particle velocity, the spectral features like metal emission lines could include a non-thermal broadening due to peculiar non--thermal motions of the gas particles of the ICM and the velocity--broadened \apec{} is a valid choice to take this into account. In the two cases explored with the $\wabs{} \times (\bapec{}_1 + \bapec{}_2 + \bapec{}_3 + \bapec{}_4 + \bapec{}_5)$ model, every frozen parameter was assumed in the same manner as in \sec\ref{sec:xspec} and the five \bapec{} normalizations were let be free. In addition, in the first model (\bapec--1), we fit for another free parameter, which is the velocity dispersion $\sigma_v$ linked across the five components, whereas in the second case (\bapec--2) both $\sigma_v$ and the redshift are left free in the fit. As shown in the plots, the results for the normalizations are definitely consistent with the standard case, where the five thermal components are described by the \apec{} model. However, the best--fit values for $\sigma_v,$ in particular, turn out to have very large errors, meaning that this additional parameter is not reliably fitted. We interpret this as an evidence that the energy resolution of the {\it Suzaku} XIS spectrometer is not sufficiently high to capture the details of the emission lines and therefore a velocity--broadened model like \bapec{} is not improving the fit to the synthetic data.

Moreover, the results look generally stable against metallicity, which could in principle compete with normalization in order to reproduce the spectral features associated to metal emission lines. In particular, we tested the fit with five \apec{} models, both for \leiaD{} and \leiaB{}, allowing for the metallicity value to be free, although linked across the five components, in addition to the five normalizations. Generally, this converges to values within few percents from the expected metallicity of $Z = 0.3\zsun,$ used in the simulation of the X--ray observation. Comparing to the results of the standard fit reported in \tab\ref{table:xspecfit}, the \apec{} best--fit normalizations, for initial metallicity values set to $[0.2,0.3,0.5]\zsun,$ are found to be consistent with the original values, within the error bars associated to them.
\section{DISCUSSION AND CONCLUSIONS} \label{sec:conclusion}
In this paper, we have presented a novel X--ray photon simulator, named \name{}\footnote{The source code of the presented version of \name{} is made available by request to the authors (see http://www.mpa-garching.mpg.de/~kdolag/Phox/). We also plan to make the code available directly for anonymous download in the near future.}, dedicated to obtain synthetic X--ray observations from the output of hydrodynamical numerical simulations.

The development of \name{} has been strongly motivated by the ultimate aim of comparing the output of hydrodynamical simulations of cluster--like objects to real X--ray observations in the most faithful and reliable way possible. In fact, X--ray observations still provide us with increasingly detailed data that require a deep understanding of the underlying structure of galaxy clusters. Numerical simulations, on the other side, continuously improve in describing not only the dark matter but also the complex physical processes governing the baryonic component. Therefore, a strategy devoted to combine the advances and results in both observations and simulations can be promisingly successful.

 In the past, software packages like X--MAS \cite[][]{gardini2004} or XIM \cite[][]{xim2009} have been developed to produce synthetic observations of astrophysical sources simulated with particle-- and grid--based codes, respectively. The existence of such tools is definitely complementary to our simulator in order to test the robustness of comparison studies between state--of--the--art hydrodynamical simulations including different physical phenomena and real observations. With respect to other codes, the novelty of our virtual telescope resides in the original method implemented. In particular, the order of the three main units that constitute the code offers a significant gain in computational cost and time, 
making \name{} extremely flexible and useful to process hydro--numerical simulations of several astrophysical objects, at various spatial scales, as well as simulated light--cones. 
Moreover, this method uniquely guarantees extremely high spatial and energy resolution. 
The more demanding phase is indeed the first one, where the output of a hydrodynamical simulation is processed and the X--ray emission of each 
gas element (either particle or grid cell) 
is simulated by using the spectral models available in the X--ray software \xspec{}. 
Essentially, we perform 
the sampling of each spectrum with an expected number of photons,
and we immediately store the sample of photons insted of constructing 
a library of template spectra to be populated afterwards.
As such, the spectral model chosen to mock up the emission from the gas elements, regardless of its complexity, can be explicitly specified and suited to represent any desired astrophysical source. Also, the relatively reasonable requirements in terms of computational time and memory, make out of \name{} an ideal tool to process forthcoming large cosmological, hydrodynamical simulations as well as the usual smaller ones.

 Concerning galaxy clusters, \name{} currently uses two possible models for the X--ray emission of the ICM (namely, \mekal{} and \apec{}) and accounts for the particular density and temperature of the simulated gas elements. However, the technique implemented offers the chance to account also for different metallicities, or even metal compositions, across the emitting gas elements, without increasing the memory requirements significantly. 

Furthermore, \name{} postpones the projection along the line of sight and the convolution with a real instrument response till after the generation of the discrete sample of photons, permitting to each of them, in principle, to contribute to the final integrated spectrum. Requiring a high energy resolution in the model spectrum associated to every gas element, we can ideally combine the potential of \name{} with the characteristics of upcoming X--ray satellites, like IXO/ATHENA or ASTRO--H, in order to predict the detectability of fundamental spectral features. This could be of great help, for instance, in constraining the ICM non--thermal motions (e.g. turbulent, bulk, rotational motions) via the velocity broadening of heavy--ion emission lines.
In general, the theoretical modelling of X--ray--emitting sources is a powerful way to predict and optimize the capabilities of future X--ray missions and to help in the interpretation of the data when the instruments are in operation.

As an interesting application of this new virtual telescope, we focused on the hydro--numerical simulation of a large--scale region, originally performed with the TreePM/SPH code GADGET--2 and then re--simulated including cooling, star formation and feedback from supernova winds. 
This filament--like structure is particularly worthy for presenting the simulator in each part and especially for demonstrating the advantages of our approach. 
From a purely technical perspective, the greatest gain was achieved in terms of computational cost since the entire, large data cube contained in the simulation output was processed with \name{} Unit 1 only once, providing an output of relatively compacted photon data (see \tab\ref{table:timing}).
In fact, this first phase permitted to obtain the X--ray photons virtually emitted by all the gas particles in the filament. 
At a later stage, we restricted our analysis to two cluster--like haloes with different thermodynamical structure, \leiaD{} and \leiaB{}, which were geometrically selected from two different regions of the simulation box. 
Conveniently, the projection and observation units of \name{} could be run separately for each halo processing the same output generated with Unit 1 and we finally obtained synthetic {\it Suzaku} images and spectra (\fig\ref{fig:xisimg}). 
The convolution with the response of the X--ray Imaging Spectrometer (XIS) on board {\it Suzaku}, via the public tool \xissim{}, provided data that could be analysed with the standard procedures of X--ray analysis. 
An important aim of this application was the reconstruction of the intrinsic emission measure distribution by the multi--temperature fitting of the simulated spectra. The goal of setting spectroscopic constraints on the multi--phase ICM of galaxy clusters was pursued in X--ray observational studies reported by \cite{peterson2003}, \cite{kaastra2004} and \cite{fabian2006}, although for real clusters it is intrinsically impossible to know the underlying multi--phase structure unambiguously. 
Instead, numerical simulations uniquely allow to model and predict the theoretical temperature distribution of the ICM \cite[e.g., studied in][]{kawahara2007,kawahara2008} and constitute therefore the ideal possibility to test the 
power of the method used to reconstruct the structure of the mock data. 
The direct comparison between the X--ray data and the theoretical thermal distribution of the input simulation is found to be overall faithful and robust for both \leiaD{} and \leiaB{} (see \fig\ref{fig:leiaDBem}).  
Moreover, the different thermal structure that distinguishes the two selected haloes is remarkably highlighted by the multi--temperature fitting.
This result promisingly suggests that the temperature distribution of the ICM can be in principle traced and unveiled by spectral analysis of X--ray observations.
From the additional investigation 
of velocity--broadened fitting models (like the \bapec{} model) for the {\it Suzaku} spectra, we can also conclude that a higher spectral resolution is required in order to reliably constrain the ICM velocity structure. In general, it is important to notice that the application discussed here does not include other emission components (e.g. a physical X--ray background) in addition to the ICM one and represents therefore an ideal case study.

With regard to technique, the main feature distinguishing \name{} from other virtual X-ray telescopes, such as XMAS  
and XIM, is the explicit creation and storage of a 3--dimensional
box of photons from the raw simulation data cube itself. 
This step is performed {\it before} and without any reference to the 
observation process and its result is stored as a separate data product, which
represents the ideal X--ray emission from all the gas volumes in the simulated region,
allowing us to employ Unit 1 of \name{} to create a photon database associated to the simulation.
In Unit 2 this photon cube is projected onto a particular observational plane and 
the photons are sampled according to the chosen aperture and observation time, as described in \eq\ref{fak}.
With the photon box pre--calculated, this step is computationally trivial and can be performed very efficiently.
As such, it could in fact be made available through an on line web service.
The result of this step is a photon list and is processed in this paper by means of the {\it Suzaku} simulator \xissim{}. Similarly, we believe it should be possible for most such virtual telescopes to be adapted to the same format
without undue difficulties. 

This leads us to an interesting modular design, with a photon simulator producing a list
of photons directly from a hydrodynamical simulation, and a virtual telescope receiving this list,
and convolving it with the characteristics of a real X--ray instrument to produce a photon event file that can
be directly compared in all its aspects to the analogous files resulting from real observations.
This design presents a proper separation of concerns between theorists and observers, where each module is clearly independent from the others and can therefore be implemented and used separately.
In a larger perspective, the well specified model for the data shared and exchanged between these units could represent 
a first step towards a standardised, interoperable way to publish
simulation results in a way that is suitable for investigation by observers. 
In particular, this is very suitable for the kind of approach pursued by the International Virtual Observatory Alliance 
(IVOA\footnote{See http://www.ivoa.net.}), and we indeed plan to deploy a web service that provides access
to the results discussed in this paper in that context\footnote{Please check http://www.mpa-garching.mpg.de/HydroSims for updates on this web service.}.
%
\section*{Acknowledgments} 
The authors gratefully thank Eugene Churazov, Hajime Kawahara, Stefano Ettori, Elena Rasia, Gayoung Chon and Umberto Maio for useful discussions, that helped improving this work. V.B. acknowledges support from and participation in the International Max Planck Research School on Astrophysics at the Ludwig Maximilian University. K.D. acknowledges the support by the DFG Priority Programme 1177 and additional support by the DFG Cluster of Excellence "Origin and Structure of the Universe". The work of G.L. was supported by Advanced Grant 246797 "GALFORMOD" from the European Research Council.
\bibliographystyle{mn2e}
\bibliography{bibl.bib}

\bsp

\label{lastpage}

\end{document}